\documentclass[conference]{IEEEtran}
\usepackage{tikz}	
\usepackage{amsmath}
\usepackage{enumerate}
\usepackage{filecontents}
\usepackage[font={bf},textfont=bf]{caption}
\usepackage{hyperref}
\usepackage{graphicx} 
\usepackage{float} 
\usepackage{subfigure} 
\usepackage{threeparttable} 
\usepackage{url}
\usepackage{tabularx}
\usepackage{subfigure}
 \usepackage[utf8]{inputenc}
\usepackage{color}
\usepackage{marvosym}
\usepackage{rotating} 
\usepackage[capitalize]{cleveref} 
\usepackage{fontawesome} 

\usepackage{wasysym} 
\usepackage[font=footnotesize]{caption}

\usepackage{soul} 
\usepackage{tcolorbox}
\usepackage[ruled,linesnumbered]{algorithm2e}
\usepackage{amsfonts,amssymb} 
\usepackage{verbatim}
\usepackage{multirow}
\usepackage{bm} 

\usepackage{array} 

\makeatletter

\usepackage{booktabs}
\newcommand{\alias}{\texttt{PhantomLiDAR}\xspace} 

\begin{document}

\title{\alias: Cross-modality Signal Injection Attacks against LiDAR}

\author{
\IEEEauthorblockN{Zizhi Jin, Qinhong Jiang, Xuancun Lu, Chen Yan, Xiaoyu Ji, Wenyuan Xu\textsuperscript{\Letter}\IEEEauthorrefmark{1} \thanks{\IEEEauthorrefmark{1} Wenyuan Xu is the corresponding author}}
\IEEEauthorblockA{Zhejiang University}
\IEEEauthorblockA{
\{\href{mailto:zizhi@zju.edu.cn}{zizhi},
\href{mailto:qhjiang@zju.edu.cn}{qhjiang},
\href{mailto:xuancun_lu@zju.edu.cn}{xuancun\_lu},
\href{mailto:yanchen@zju.edu.cn}{yanchen},
\href{mailto:xji@zju.edu.cn}{xji},
\href{mailto:wyxu@zju.edu.cn}{wyxu}\}@zju.edu.cn}
}

\IEEEoverridecommandlockouts
\makeatletter\def\@IEEEpubidpullup{6.5\baselineskip}\makeatother
\IEEEpubid{\parbox{\columnwidth}{
    Network and Distributed System Security (NDSS) Symposium 2025\\
    23 - 28 February 2025, San Diego, CA, USA\\
    ISBN 979-8-9894372-8-3\\
    https://dx.doi.org/10.14722/ndss.2025.23997\\
    www.ndss-symposium.org
}
\hspace{\columnsep}\makebox[\columnwidth]{}}

\maketitle

\begin{abstract}
LiDAR (Light Detection and Ranging) is a pivotal sensor for autonomous driving, offering precise 3D spatial information. 
Previous signal attacks against LiDAR systems mainly exploit laser signals. In this paper, we investigate the possibility of cross-modality signal injection attacks, i.e., injecting intentional electromagnetic interference (IEMI) to manipulate LiDAR output. Our insight is that the internal modules of a LiDAR, i.e., the laser receiving circuit, the monitoring sensors, and the beam-steering modules, even with strict electromagnetic compatibility (EMC) testing, can still couple with the IEMI attack signals and result in the malfunction of LiDAR systems. Based on the above attack surfaces, we propose the \alias attack, which manipulates LiDAR output in terms of \textit{Points Interference}, \textit{Points Injection}, \textit{Points Removal}, and even \textit{LiDAR Power-Off}. 
We evaluate and demonstrate the effectiveness of \alias with both simulated and real-world experiments on five COTS LiDAR systems. 
We also conduct feasibility experiments in real-world moving scenarios.
We provide potential defense measures that can be implemented at both the sensor level and the vehicle system level to mitigate the risks associated with IEMI attacks. Video demonstrations can be viewed at \textcolor{blue}{\href{https://sites.google.com/view/phantomlidar}{https://sites.google.com/view/phantomlidar}}.

\end{abstract}


\section{Introduction}
\label{intro}
Autonomous systems empowered by artificial intelligence are having a transformative impact on our society. LiDAR (Light Detection and Ranging), which directly measures the coordinates and shapes of objects with precision, has been increasingly integrated into various applications such as autonomous vehicles (AVs), drones, and robots. 
According to Yole~\cite{yole2023LiDAR}, more than 119 car models are to be released with LiDAR by OEMs all over the world in 2023 or shortly thereafter. 

LiDAR plays a critical role in autonomous vehicles to perceive surrounding environments and make intelligent decisions. The safety and reliability of autonomous vehicles highly depend on the trustworthiness of the LiDAR perception.
A critical question is how safe LiDAR systems are facing surrounding physical signals. Numerous studies~\cite{petit2015remote,shin2017illusion,cao2019adversarial,jin2022pla,hallyburton2022security,cao2023you} indicate that LiDAR can be compromised by lasers. However, with lasers operating on the same physical channel, attacks typically target the photoelectric sensor in the ToF circuit, and the attack principle focuses on the transduction process. 
To explore a more expansive set of attack vectors besides the laser signals, we propose and investigate the possibility of the \textbf{cross-modality signal attacks}. A recent work~\cite{bhupathiraju2023emi} demonstrates that EMI can
 compromise LiDAR’s ToF circuits and induce attack effects of "sensor data perturbations". Overall, all previous works compromise LiDAR  by attacking the circuits or photoelectric sensor in ToF module. Whether there are new vulnerabilities and new attack surfaces against lidar systems remains an open question.

\begin{figure}[pt]
    \centering

    \includegraphics[width=0.48\textwidth]{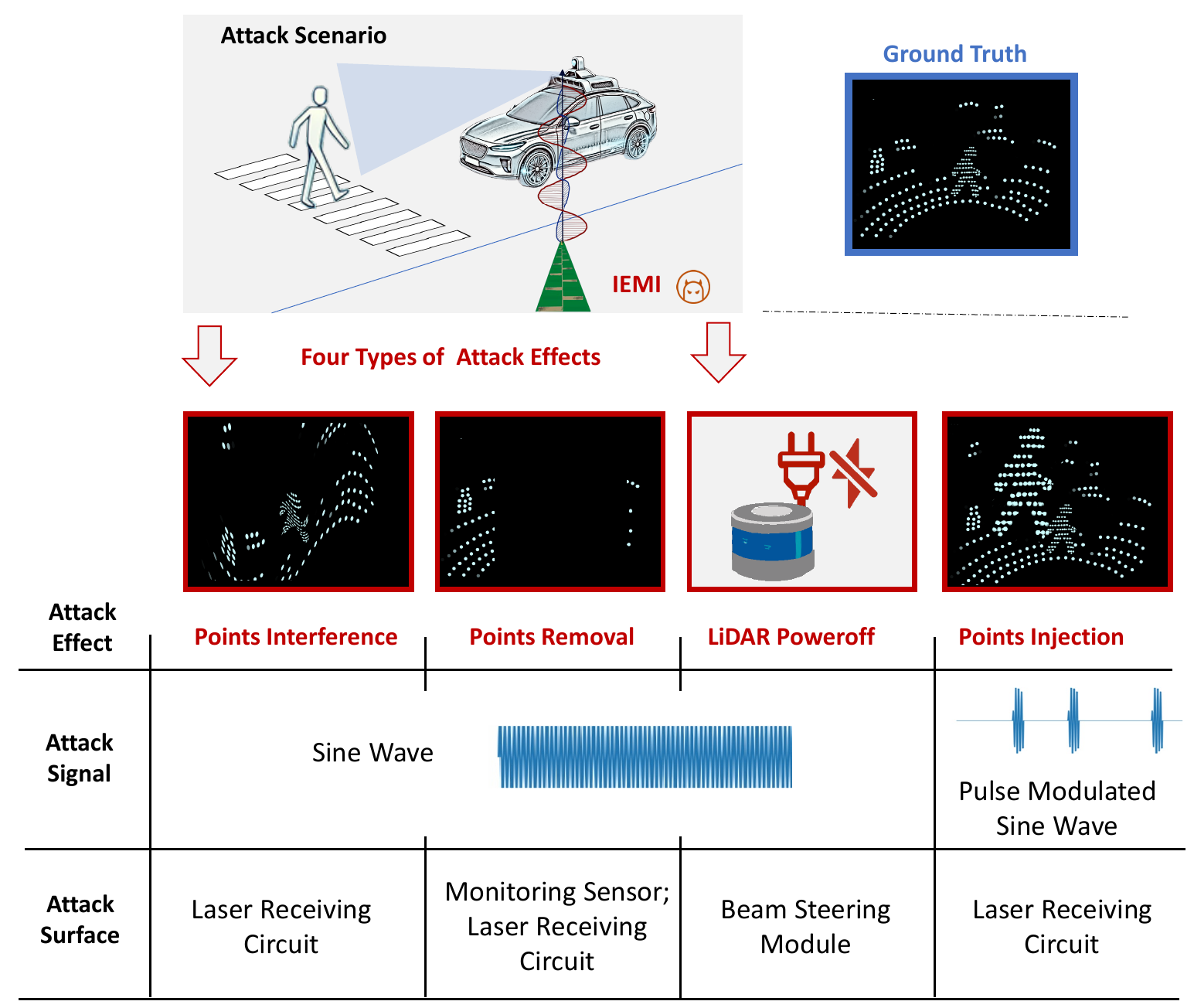}
    \vspace{-0.2in}
    \caption{Illustration of the \alias attack. \textmd{By injecting different signals into diverse attack surfaces, i.e., the laser receiving circuit, the monitoring sensor, and the beam steering module, \alias can succeed in achieving the \textit{ Points Interference, Points Removal, Points injection}, and even \textit{LiDAR Power-Off} attacks.}}
    
    \vspace{-10pt}
     \label{fig:intro}
    
\end{figure}




In this paper, our goal is to explore novel attack vulnerabilities and new attack vectors for compromising LiDAR using IEMI. 
The challenge of achieving this goal is considerable for two reasons: (1) Exploiting novel attack vulnerabilities proves difficult owing to the robust encapsulation and tamper-proof features of commercial off-the-shelf (COTS) LiDAR systems, which conceal their internal mechanics and complicate the analysis of underlying principles. (2) LiDAR systems typically exhibit resistance to IEMI due to rigorous Electromagnetic Compatibility (EMC) testing and the incorporation of anti-interference designs, such as shielding and layout optimization, which mitigate IEMI effects.

To tackle these challenges, we systematically analyze the LiDAR's internal architectures by referring to reverse engineering reports~\cite{vlp16_teardown,RSM1_teardown} and infer that the monitoring sensors (e.g., temperature sensors~\cite{tu2019trick},hall effect sensors~\cite{wang2023volttack}) and the beam-steering module can be good candidates to couple with EMI signals. 
Then, we establish a set of high-performance EM attack devices with a broad frequency range to experimentally search for vulnerabilities. Subsequently, we identify attack surfaces by utilizing the fault detection and diagnostic (FDD) mechanism and conducting validation experiments on LiDAR's internal circuits. 
Through these steps, we discovered two new attack surfaces: monitoring sensors (temperature sensors and hall effect sensors) and the optical encoder in beam steering module. Based on this, we can achieve novel attack effects such as points removal and LiDAR power-off. 


Furthermore, to explore the feasibility of skillfully manipulating LiDAR systems using EM signals for precise control, we experiment with injecting controllable points into the LiDAR system. Previous efforts~\cite{petit2015remote,shin2017illusion,cao2019adversarial,jin2022pla,hallyburton2022security,cao2023you} in this domain typically involved using lasers which forge LiDAR echoes to inject controllable points. However, directly employing the same method with EM signals to forge LiDAR echoes faces the challenge of signal coupling into the circuit. To overcome this challenge, we utilize amplitude modulation (AM). We identified an appropriate carrier frequency and modulated the fine-grained baseband signal onto the sinusoidal carrier signal, thereby achieving effective signal injection. Through this method, we successfully managed to inject controllable points using EM signals. In this paper, only the points injection attack requires signal amplitude modulation.

In general, as illustrated in Fig.~\ref{fig:intro}, we successfully implemented four types of attacks: 1) \textbf{Points Interference.} By injecting EMI into laser receiving ciruit, we can introduce errors (up to 10cm in this paper) in LiDAR ranging, thereby distorting the point cloud. 2) \textbf{Points Removal:} By injecting EMI into monitoring sensors or laser receiving ciruit, this attack causes the point cloud to deviate significantly from its true position or to disappear completely. The effect can be applied to part or all of the point cloud. 3) \textbf{LiDAR Power-off:} By injecting EMI into beam steeing module, this attack causes the LiDAR system to shut down and stop working. Even after the attack stops, the LiDAR system must be manually rebooted before it can resume operation. 4) \textbf{Points Injection:} By injecting amplitude-modulated EMI into laser receiving ciruit, this attack allows for the injection of controllable points.

Among these, \textit{Points Removal}, \textit{LiDAR Power-off} and \textit{Points Injection} are three new attacks. Additionally, the \textit{Points Interference} attack enhances the attack capability established in previous work~\cite{bhupathiraju2023emi}. We hope that the increased types of attack effects and enhanced attack capabilities can help the security community and LiDAR manufacturers accurately recognize the threats posed by EMI to LiDAR systems. This recognition should further promote the establishment of more advanced EMC standards and the design of more secure LiDAR systems.

To evaluate the PhantomLiDAR attacks, we explore the IEMI vulnerability on five COTS LiDAR systems including three rotating LiDARs and two MEMS LiDARs. To better understand the advantages and limitations of the four types of attacks, we designed unique evaluation methods for each. We evaluate the attack in both emulated and real-world setups on five 3D object detection models, considering the impacts of attackers' location and aiming. Notably, we validate that PhantomLiDAR can hide a targeted object even when the attack distance is 5 meters away. In addition, PhantomLiDAR can inject over 16,000 fake points in the VLP-16 LiDAR, a quantity that is five times greater than that achieved by SOTA laser-based attacks~\cite{jin2022pla}, which could inject less than 3,000 fake points in the VLP-16 under the same rotation speed. What's more, we conduct feasibility experiments in moving scenarios.

We summarize our main contributions as follows: 
\begin{itemize}
    \item \textbf{Attack Surfaces:} As far as we know, we are the first to propose the attack surfaces of monitoring sensors  and optical encoder in beam-steering module on LiDAR.

    \item \textbf{Attack Effects:} We propose three new attack effects including  \textit{Points Removal}, \textit{LiDAR Power-off}, and \textit{Points Injection}.
     

    \item \textbf{Attack Capabilities:} The \textit{Points Interference} show 2x stronger interference capability compared to SOTA works. The \textit{Points Removal} can hide a target remotely without precise aiming. The \textit{LiDAR Power-off} can success on popular mehchanical LiDAR VLP-16 and MEMS LiDAR RS-M1. \textit{Points Injection} can inject controllable points number 5x more than SOTA laser-based attacks.
    
    \item \textbf{Experiments:} We conducted experiments on five COTS LiDAR systems. For the four types of attacks, we designed specific simulated and real-world experiments to better evaluate their effectiveness and limitations.

\end{itemize}

\section{Background}
\label{sec:background}

\subsection{LiDAR System}
\label{sec:lidarsystembackground}

\begin{figure}[pt]
    \centering

    \includegraphics[width=0.48\textwidth]{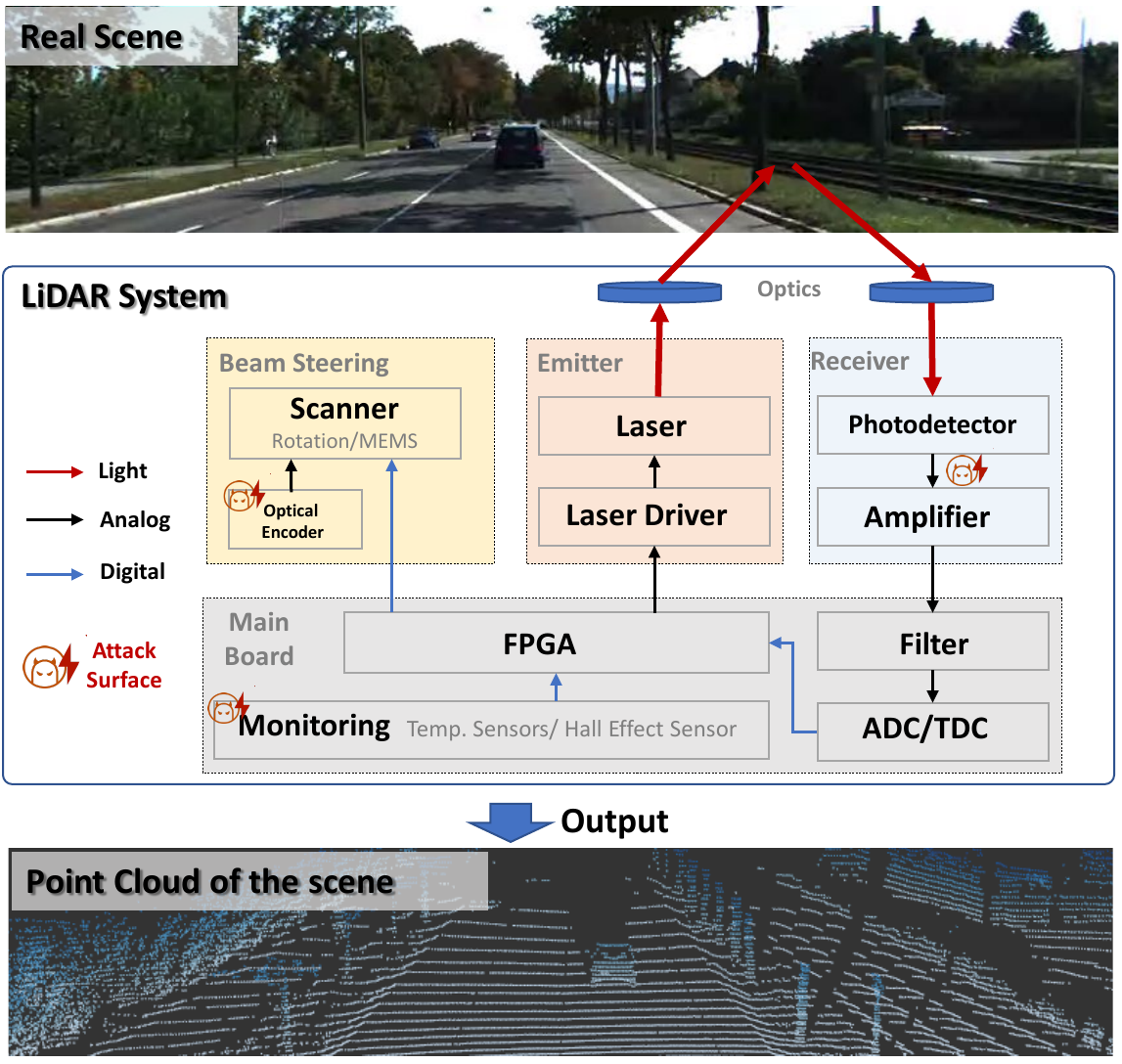}
    \vspace{-0.2in}
    \caption{LiDAR System. \textmd{A typitcal LiDAR system includes one (multiple) emitter-receiver pair (pairs) for ranging, a beam-steering module for laser scanning and a main board for controlling.}}
    \vspace{-10pt}
     \label{fig:lidarsystem}
    
\end{figure}

LiDAR provides precise 3D spatial information through point cloud data.
Fig.~\ref{fig:lidarsystem} illustrates the main components of a typical lidar, which include an emitter-receiver pair (or pairs). 
During a ranging process, the main board controls the emitter to emit the laser signal, and its direction and firing time $\tau_{0}$ are registered. 
The laser pulse travels through air, and when it hits an object, a portion of that energy is reflected and received by the paired receiver. The light signal is then converted into an analog electrical signal through a photoelectric sensor. 
After passing through an amplifier, filter, and ADC, the analog signal is transformed into a digital signal that is input into the FPGA. 
The algorithms within the FPGA can determine the receiving time $\tau_{1}$ and intensity of the echo signal.
The range $R$ is measured based on the round-trip delay of light to the target:
\begin{equation}
   R =  \frac{1}{2} c\cdot (\tau_{1}-\tau_{0}) 
\end{equation}
where $c$ is the speed of light in the medium (e.g., air) between the LiDAR and the target. This process is called Time-of-Flight (ToF) ranging. Based on this equation, the lidar-based range measurement is equivalent to measuring the round-trip delay of light waves to the target. This is achieved by modulating the intensity, phase, and/or frequency of the waveform of the transmitted light and measuring the time required for that modulation pattern to appear back at the receiver. 

To create the point cloud, the light should be directed to all the points in a desired field of view (FOV). This can be done by employing a beam-steering unit to scan the FOV. Over the years, many different beam-steering techniques have been developed. Foremost among these are mechanical motion of the light source~\cite{velodyneVLP16,robosenseRS16,halterman2010velodyne}; deflection of the light using a macro~\cite{innovusionFalcon} or micro-mechanical mirror~\cite{robosenseM1}; optical-phased arrays~\cite{zhang2022large}.
To date most commercially available LiDAR system have been direct detection time-of-flight (ToF) sensors operating at 905 nm using mechanical motion or mirrors for beam steering~\cite{yole2023LiDAR,shi2019photonic}. In our study, we primarily focus on these commercially available LiDAR systems, as they are broadly utilized in today's autonomous vehicles. There is usually an optical encoder in the beam steering module to monitor the running state of the motor.

In addition to the aforementioned components related to point cloud generation, LiDAR manufacturers typically incorporate state monitoring sensors to oversee the operational status of these modules, such as the state of supply voltage (hall effects sensor), temperature.  
In this paper, we experimentally validate that the receiving circuits, the monitoring sensors and the optical encoder can be the attack surfaces for IEMI.
\subsection{LiDAR Fault Detection and Diagnostic}
\label{sec:lidarFDDbackground}
To ensure that the LiDAR operates as intended, the LiDAR manufacturers typically utilizes fault detection and diagnostic (FDD)~\cite{abid2021review,goelles2020fault} mechanisms to ensure these modules function as intended. 
If a LiDAR system fails to operate as expected, it could potentially cause damage. For instance, excessive laser emission power or a motor malfunction leading to continuous laser exposure on a single spot could pose risks to the eyes. Therefore, for the protection of both people and the LiDAR system itself, FDD and response mechanisms are often integrated into the design of the LiDAR.  

\begin{figure}[pt]
    \centering

    \includegraphics[width=0.48\textwidth]{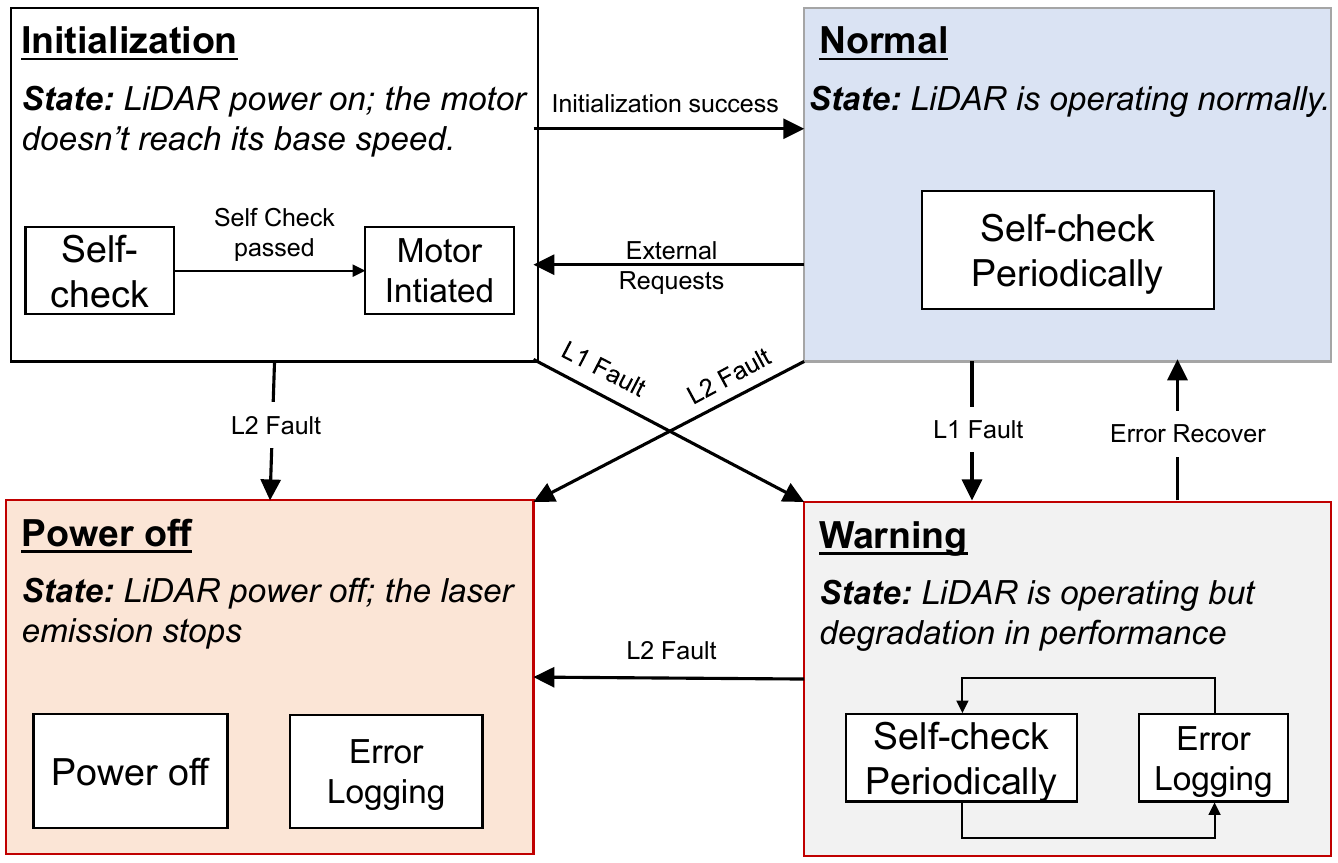}
    \vspace{-0.1in}
    \caption{LiDAR fault detection and diagnostic. \textmd{There are four typical states during LiDAR operation: Initialization, Normal, Warning, and Power-off. A LiDAR system may alternate among the four operational states when different-level faults are detected.}}
    \vspace{-10pt}
     \label{fig:fault_diagnostic_management}
    
\end{figure}

According to the patents released by LiDAR and car manufactures~\cite{HesaiLidarPatent,RobosenseLidarPatent,GMLidarPatent}, manufacturers categorize common LiDAR faults and classify them based on severity and consequences into two levels: Level 1 (L1) faults are those with lower impact. The LiDAR can continue to operate under L1 faults but with reduced performance or parameters; Level 2 (L2) faults are more severe, which either render the LiDAR inoperable or lead to performance degradation beyond acceptable limits. The LiDAR may shut down when the L2 faults are diagnosed.
As shown in Fig.~\ref{fig:fault_diagnostic_management},  there are four typical states during LiDAR operation: \textit{Initialization, Normal, Warning, and Power-off}. A LiDAR system may alternate among the four operational states when faults are detected. When a LiDAR powers on, it first enters \textit{Initialization} and performs self-check. Then the LiDAR activates the motor, and once the motor speed has reached the preset value and the self-check is passed, it enters the \textit{ Normal} operating state. In the \textit{ Normal} state, the LiDAR periodically performs self-check. If an L1 fault is detected during the \textit{Initialization} or \textit{Normal} states, the LiDAR enters the \textit{Warning} state.  If an L2 fault is detected at any state, the LiDAR will enter the \textit{Power-off} state, where it typically shuts down power or communication. 

A typical FDD list from an anonymous LiDAR manufacturer is shown in Table.~\ref{table:Fault_Matrix} in the Appendix. It should be noted that different LiDAR manufacturers may have their own definitions for diagnosing and managing faults in LiDAR systems.

 \subsection{Backgroud of IEMI Attacks}

The IEMI term is officially defined by the International Electrotechnical Committee (IEC) as ``intentional malicious generation of electromagnetic energy introducing noise or signals into electric and electronic systems, thus disrupting, confusing, or damaging these systems for terrorist or criminal purposes''~\cite{radasky2010intentional}. 
The adversary should first consider possible ways to couple the IEMI signals into the target to implement a successful IEMI attack. We refer to this process as constructing the coupling channel, and there are three essential components for a coupling channel:

\textbf{Attack Surfaces} is a ``wire'' that exists in the victim circuit that acts as an unintentional receiving antenna to be interfered with by the malicious IEMI signals. This kind of ``wire'' can be analog electrical traces on the printed circuit boards (PCB)~\cite{kohler2022signal,dayanikli2020electromagnetic,szakaly2023assault,jiang2024ghosttype,kune2013ghost,xu2021inaudible,fokkens2021machine} or the digital sensory communication channel between the sensor and controller~\cite{jiang2023glitchhiker,jang2023paralyzing}.

\textbf{Coupling Path} decides how the IEMI generated by attackers reaches the victim device. The selection of coupling paths largely depends on the target coupling interface to achieve the best coupling efficiency. There are two categories of coupling paths: radiated and conducted.
Radiated coupling propagates electromagnetic energy through the air or vacuum without making any physical contact with the target, such as magnetic coupling (changing magnetic fields ($H$-fields)), electrical coupling (changing electric fields ($E$-fields)), and electromagnetic coupling (changing both magnetic and electric fields ($EM$-fields)).
Conducted coupling propagates electromagnetic energy from an electromagnetic source to a coupling interface via wire conduction~\cite{zhu2022powertouch}.

\textbf{IEMI Sources.} To deliver the IEMI signals into target systems efficiently, attackers need to generate IEMI signals at frequencies that are compatible with the electrical characteristics of the target systems and at suitable amplitudes that can change system data in the desired manner. 

\section{Threat Model}
\subsection{Attack Goal}
The attacker’s goal is to diminish the reliability and performance of LiDAR systems by stealthily interfering with the point cloud or LiDAR operating status using EM signals. Specifically, we consider 4 types of attacks:
\begin{itemize}
    \item \textbf{Points Interference:} The goal of this attack is to introduce errors in the distance measurement of the LiDAR.
    \item \textbf{Points Removal:} The goal here is to significantly displace the point cloud of an object from its actual position or to completely erase it, consequently preventing the LiDAR-based perception models from detecting the object.
    \item \textbf{LiDAR Power-off:} The effect of this attack is to shut down the LiDAR, rendering it inoperative. Even after the attack stops, the LiDAR requires a reboot to resume functioning.
    \item \textbf{Points Injection:} This attack focuses on injecting false points with controllable positions and patterns.
\end{itemize}

\subsection{Attacker's Capability}

We consider the attacker with the following assumptions.

\textbf{EM Attack Capability to the LiDAR.} The attacker can transmit EM signals to the LiDAR in the victim autonomous cars or robots remotely without attaching any hardware or software on the target LiDAR systems. To achive it, we assume the attacker is equipped with commercial devices that can generate EM signals including an RF antenna, a signal generator and an RF power amplifier. 
The attack devices can be set up in the attacker's car, allowing the attacker to follow the victim vehicle and conduct EM injection attacks within a certain distance. They can also park on the roadside to attack passing vehicles or those stopped at red lights. 

\textbf{Budget-related Considerations.} Attackers may require high-end devices to perform wide-frequency range sweeps in order to identify the vulnerabilities of LiDAR systems. Once these vulnerabilities are identified, lower-cost attack devices can be used to carry out the attack.


\textbf{LiDAR Parameter Awareness.} We assume the attacker knows the model of the target LiDAR, and she may obtain a similar substitute LiDAR for assessment beforehand. For example, she may implement frequncy sweep experiments on the substitude LiDAR to find a vulnerability frequency offline before officially implement attack.

\textbf{Black Box.}  The adversary does not have access to the machine learning model or the perception system. Attackers can exploit only the characteristics and vulnerabilities of the sensors to achieve their attack target.




\section{Preliminary Study: Feasibility and Vulnerability Analysis}
\label{sec:feasibility study}

In this section, we first investigate the feasibility of efficient electromagnetic injection and its potential attack effects on LiDAR systems. Subsequently, we analyze the principle of various attack effects.


\subsection{Attack Intuition}

There are two attacks intuitions can be formulated that can potentially compromise LiDAR: direct attack and indirect attack.


\textbf{Direct Attack:} 
From Sec.~\ref{sec:lidarsystembackground}, we learned that a LiDAR point is generated by a ToF ranging process, in which a laser signal is converted into an analog electrical signal by a photodetector in the receiving circuit. Therefore, a direct way to compromise the point cloud is to interfer with the analog signal in the receiving module, directly affecting the LiDAR's ranging mechanism and subsequently disrupting the point cloud.


\textbf{Indirect Attack:} From Sec.~\ref{sec:lidarFDDbackground} and Table.~\ref{table:Fault_Matrix}, we learned that when the FDD mechanism of a LiDAR detects a fault, the LiDAR enters an abnormal self-protection status, in which it may treat the point cloud as invalid or even shut down. Therefore, it may be feasible to compromise other modules in LiDAR, e.g., temperature sensors~\cite{tu2019trick},  power supplies\cite{wang2023volttack}, and optical encoder in the beam-steering unit~\cite{shoukry2013non}. This may induce the LiDAR to detect errors, triggering the FDD’s inherent operations, forcing the LiDAR into fault recovery, and thereby leading to denial of service or even shut down.




\begin{figure}[pt]
    \centering

    \includegraphics[width=0.48\textwidth]{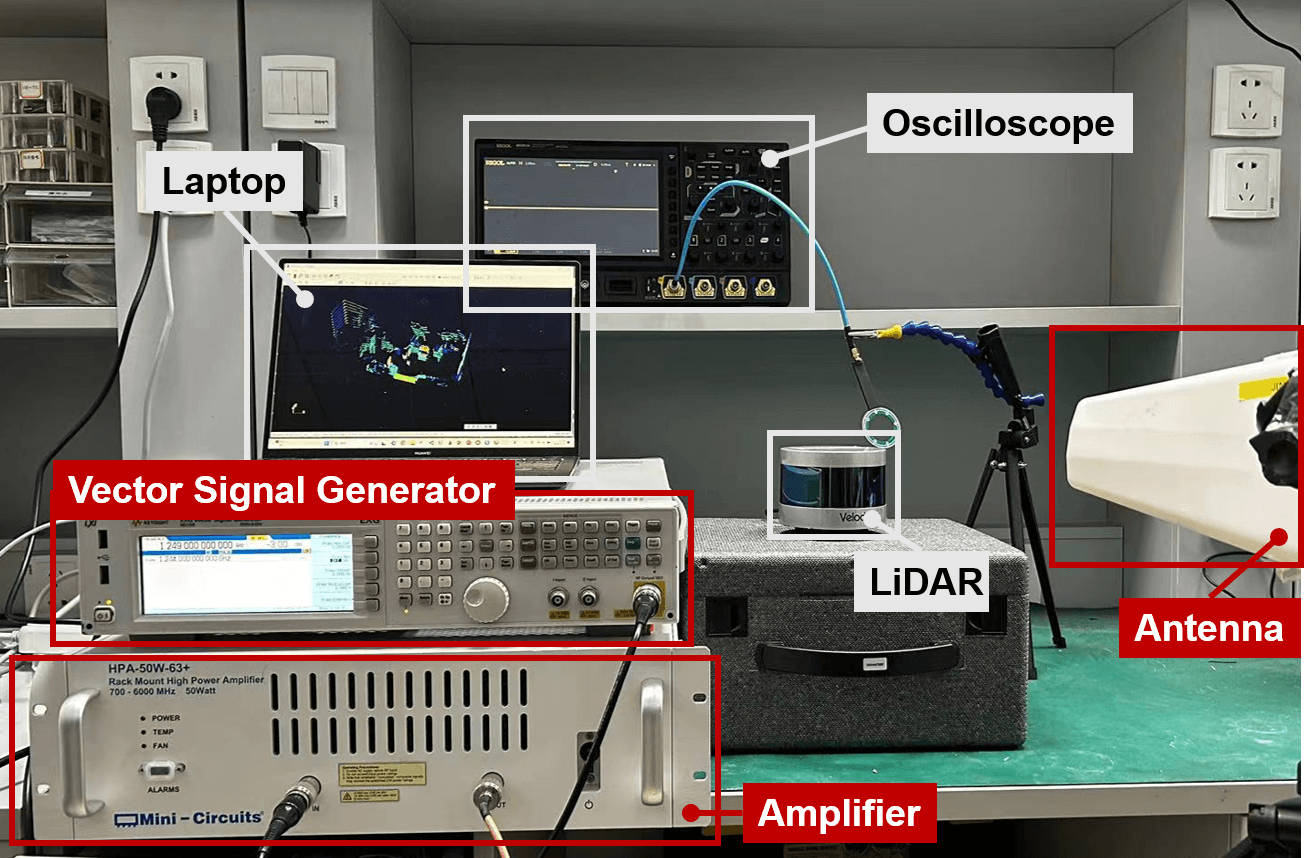}
    \vspace{-0.1in}
    \caption{The testbed for IEMI attack against LiDAR}
    \vspace{-10pt}
     \label{fig:Feasibility_Setup}
    
\end{figure}


    





\subsection{Feasibility of IEMI attack on LiDAR}


When we attempt to inject EM signals to compromise a LiDAR, we consider the LiDAR's analog electrical traces on the PCB and the electric wire as receiving antennas. 
The frequency of the EM signal determines the efficiency of its reception by a specific antenna. Theoretically, we can estimate the resonant frequency where the maximum coupling efficiency occurs based on the length and shape of the targeted/selected antenna~\cite{redoute2009emc}. However, accurately calculating the signal frequency is challenging due to the unknown nature of the target antenna in LiDAR. To address this challenge, one of the most commonly used methods is frequency sweep.

\subsubsection{Setup}
The experimental setup for the feasibility study is shown in Fig.~\ref{fig:Feasibility_Setup}. The attack devices include a Keysight N5712b vector signal generator for EMI signal generation, a Mini-Circuits HPA-50W-63+ power amplifier for amplifying the EMI signal, and a log-periodic antenna for signal transmission. The LiDARs under test is the VLP-16~\cite{velodyneVLP16}, which is the most popular LiDAR in related work~\cite{shin2017illusion,cao2019adversarial,jin2022pla,bhupathiraju2023emi}.

We conduct frequency sweep ranging from 500~MHz to 3500~MHz with an interval of 1~MHz. The signal generator output is set at 0~dbm with an amplifier gain of 50~W. The video demo of the frequency sweep can be found on the website~\cite{PhantomLiDAR}. We record the point cloud and observe the running status of LiDAR under EMI at various frequencies. Both quantitative calculation and attack effects analysis are employed to present the results of the frequency sweep.



\subsubsection{Quantitative Analysis}
\begin{figure}[tp]
	\centering
	\subfigure[\textbf{Quantitative analysis.} We utilize Hausdoff distance to quantify the distortion of point clouds under EMI at various frequencies.]{
        \includegraphics[width=0.95\linewidth]{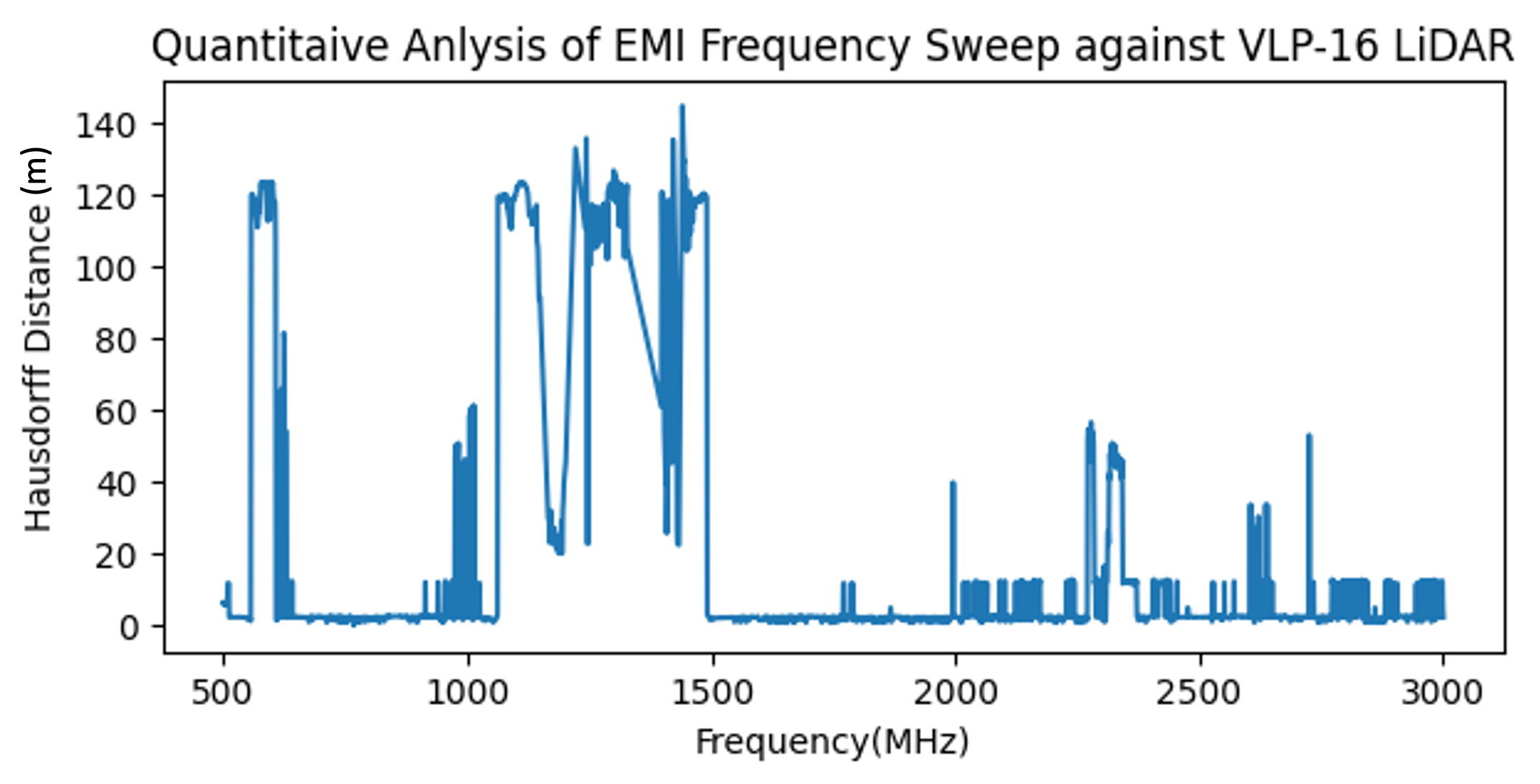}
         \vspace{-0.1in}
        \vspace{-10pt}
         \label{fig:quantitativesweepfrequency}
	}
 
	\subfigure[\textbf{Attack Effects.} By employing EM signals of varying frequency, we can induce effects such as \textit{Point Interference}, \textit{Point Removal} and \textit{LiDAR Power-off} against VLP-16 LiDAR.]{
	\includegraphics[width=0.95\linewidth]{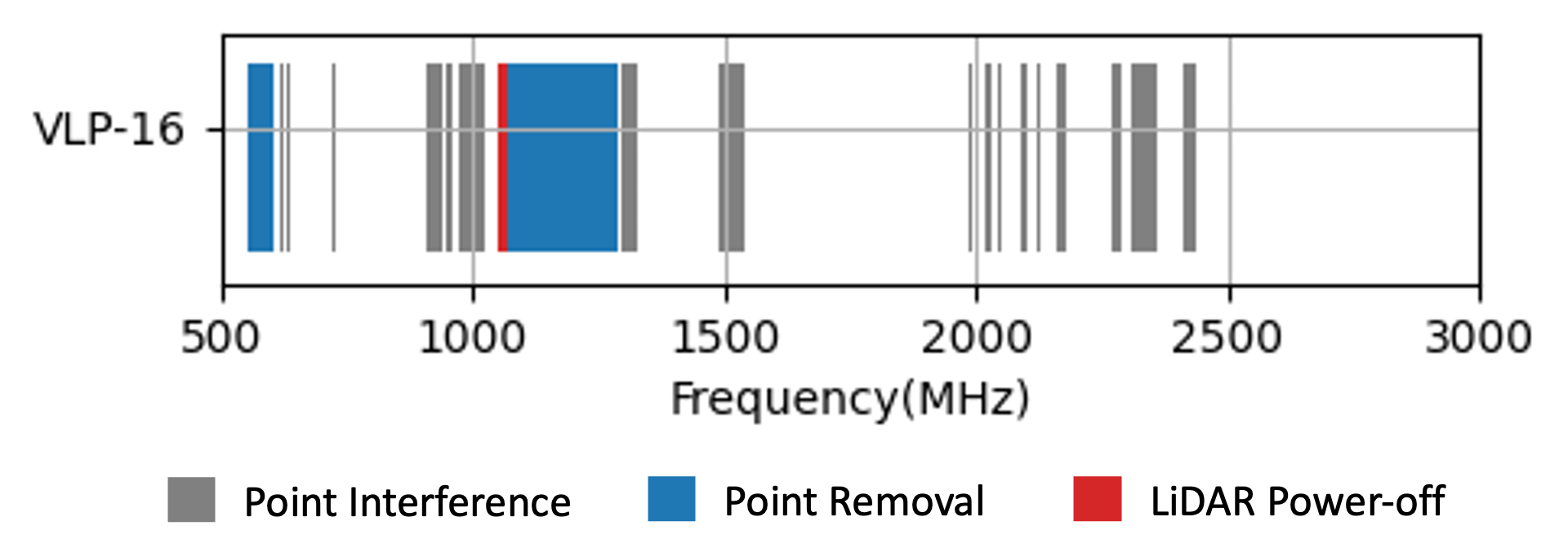}
        \vspace{-0.1in}
        \vspace{-10pt}
         \label{fig:VLP-16_Sweep_attack_effects}
     
	}

        \subfigure[\textbf{Illustration of the Attack Effects.} The \textit{Points Interference} and \textit{Points Removal} on VLP-16. Best viewed on a screen and zoomed in.]{
	\includegraphics[width=0.95\linewidth]{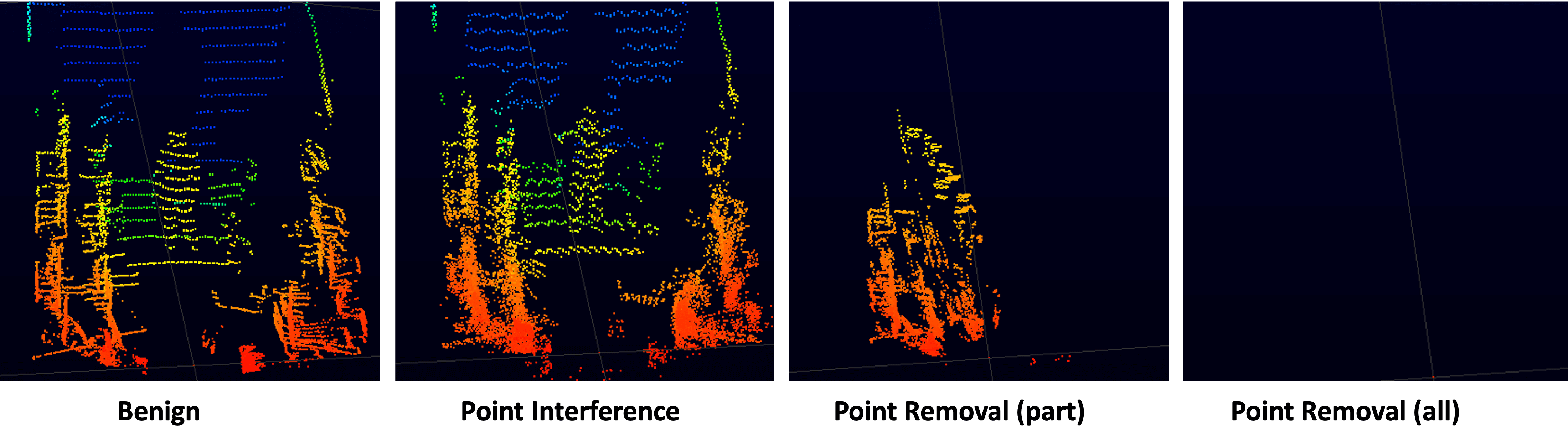}
        \vspace{-0.1in}
        \vspace{-10pt}
         \label{fig:attackeffects_interference_eliminate}
     
	}
 
        \vspace{-0.1in}
	\caption{Feasibility Experiments.}
        \label{fig:feasibility_experiments}
	\vspace{-0.1in}
        \belowdisplayskip=12pt
\end{figure}

In the quantitative analysis, we utilize the parameter \textit{Hausdoff distance}~\cite{huttenlocher1993comparing} between the benign point cloud and the interfered point cloud to quantify the level of distortion for point clouds under IEMI. 
In mathematics, the Hausdoff distance measures how far two subsets of a metric space are from each other. Consider the benign point cloud $\mathbb{PC}$ and the interfered point cloud $\mathbb{PC'}$, the Hausdorff distance between them is denoted as $D_{H}(\mathbb{PC},\mathbb{PC'})$. A larger value of $D_{H}(\mathbb{PC},\mathbb{PC'})$ indicates a stronger degree of interference by IEMI on the LiDAR system. During the sweep process, apart from the varying frequency of the EM signal, we strive to maintain a consistent laboratory environment. After recording the point clouds during frequency sweep, we calculate the Hausdorff distance between point clouds at each frequency ranging from 500~MHz to 3500~MHz and the benign point cloud. The results, as shown in Fig.~\ref{fig:quantitativesweepfrequency},  reveal that many different frequencies can cause significant interference effects. For instance, at frequencies around 1200~MHz, the Hausdorff distance reaches as high as 120 meters. Observing the point cloud under 1200~MHz EMI, we find that all points have been erased. In summary, the quantitative analysis of frequency sweep validates the feasibility of EMI attacks on LiDAR. However, to systematically investigate the principles of these attacks, we also need to focus on the effects of the attacks.

\subsubsection{Attack Effects Analysis}
From quantitative analysis, it can be observed that signals of different frequencies can cause varying degrees of interference with LiDAR point clouds. 
To systematically analyze the principles of attacks, we categorize them based on the extent of interference in the point clouds under EMI and the changes in the operational state of the LiDAR. The attack effects are intuitively categorized as \textit{Points Interference}, \textit{Points Removal}, and\textit{ LiDAR Poweroff}. The results of attack effect analysis on VLP-16 are presented in Fig.~\ref{fig:VLP-16_Sweep_attack_effects}. 

The criteria for categorising attack effects are presented below. In a static environment, we measure the Euclidean distance between points on the same ray before and after an attack. If the average Euclidean distance is less than 2~cm, we consider the LiDAR unaffected by Electromagnetic Interference (EMI), given that the inherent range accuracy of LiDAR is 2~cm. If the average Euclidean distance is greater than 2~cm but less than 1~m, we define this type of attack effect as \textit{Points Interference}. Should the average Euclidean distance exceed 1~m, it implies that the points have been displaced from their original positions, leading us to categorize this effect as \textit{Points Reomval}. \textit{Points Removal} can affect either part or all of the point cloud. Furthermore, we find that when the electromagnetic frequency is swept to $1040 \sim 1070$~MHz, the VLP-16 LiDAR system shuts down completely and fails to recover even after cessation of the EM attack. The system must be rebooted to resume operation. This type of attack effect is categorized as \textit{LiDAR Power-off}. 



    


    

\subsection{Principle Analysis and Validation}

We analyze and validate the principle of the aforementioned three types of attack effects in this section. 


\subsubsection{Why Points Interference Can be Induced}
\label{sec:Point_Interference_analysis}
\


 



    
\begin{figure}[tp]
	\centering



        \subfigure[\textbf{Principle of \textit{Point Interference}.} The sinusoidal interfering signal can be injected into the receiving circuit by EMI, causing minor variations in the peak time of the return signal. This subsequently causes a slight shift in the position of the points, either forwards or backwards. This distance shift is defined as distance error, which can quantify the intensity of \textit{Points Interference}.]{
        \includegraphics[width=0.95\linewidth]{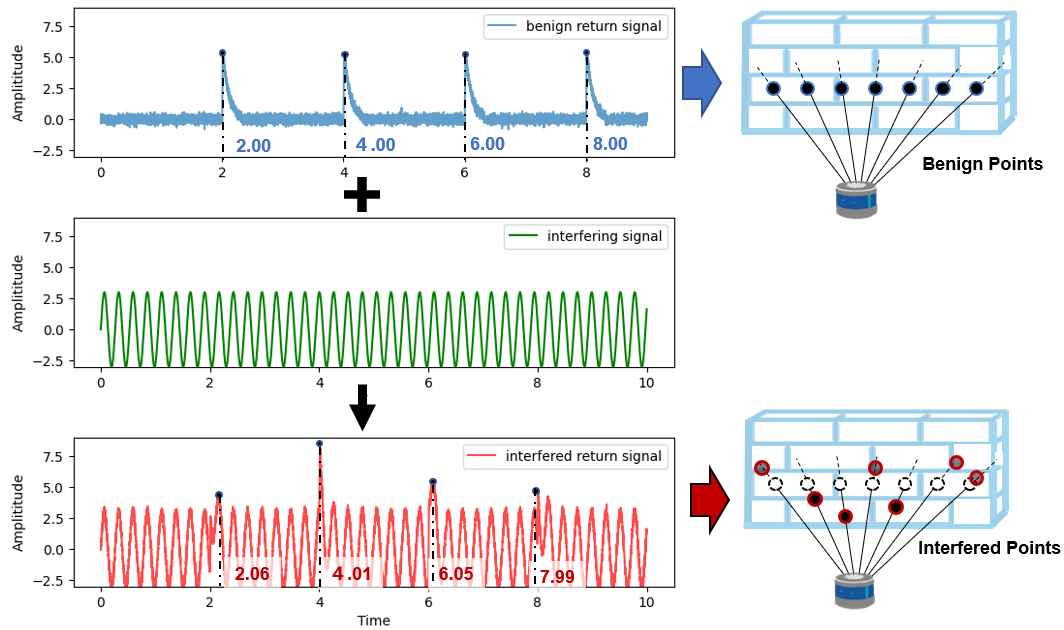}
         \vspace{-0.1in}

        \vspace{-10pt}
         \label{fig:point_interference_principle}
	}

        \subfigure[\textbf{The patterns of Points Interference.} Due to the aliasing effect caused by ADC undersampling, different frequencies can cause different interference patterns.]{
	\includegraphics[width=0.95\linewidth]{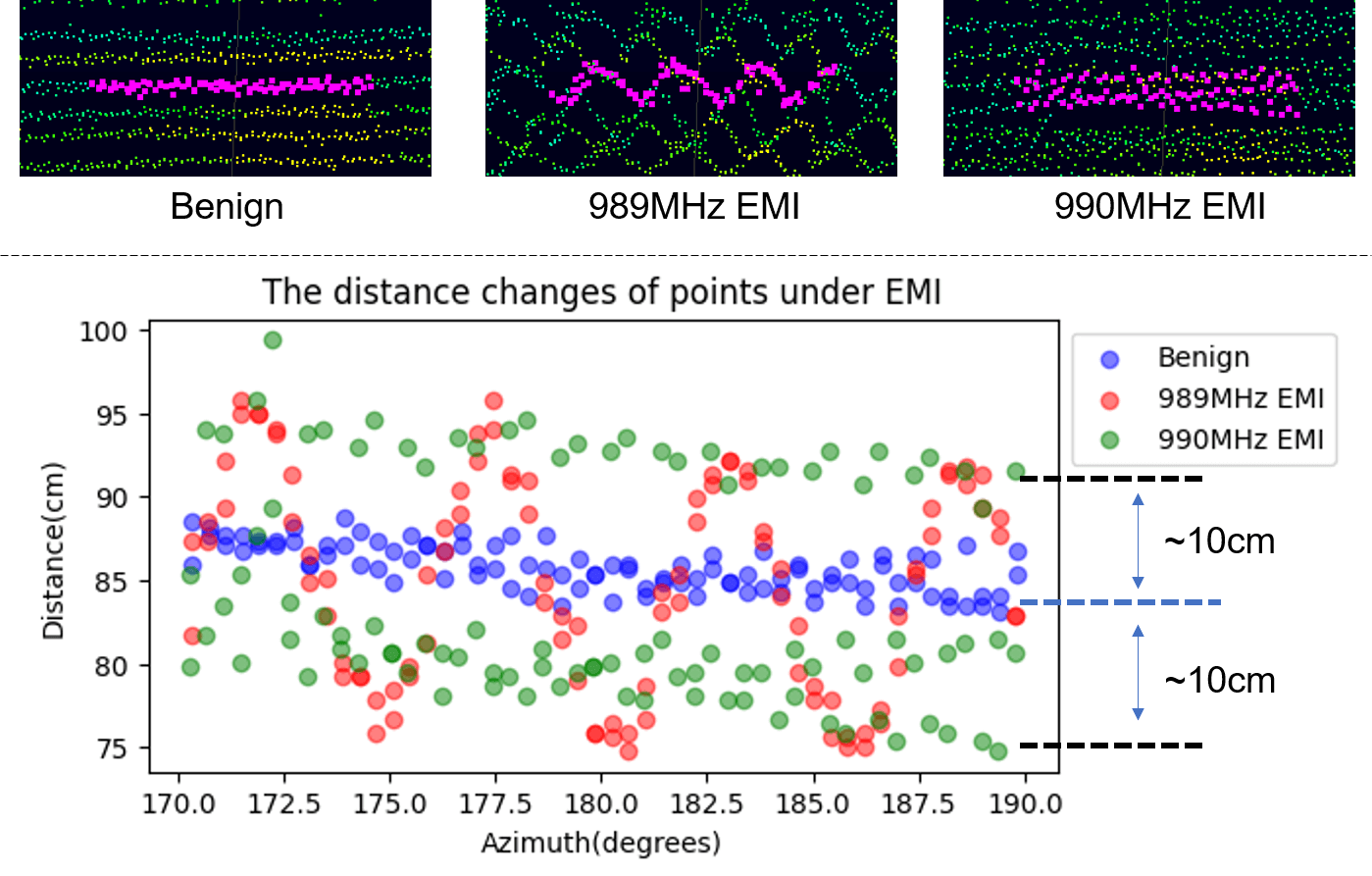}
        \vspace{-0.1in}
        \vspace{-10pt}
         \label{fig:points_interference_pattern}
	}

	\subfigure[\textbf{Principle validation.} EMI can couple into the transmission line in the receive board of LiDAR.]{
	\includegraphics[width=0.95\linewidth]{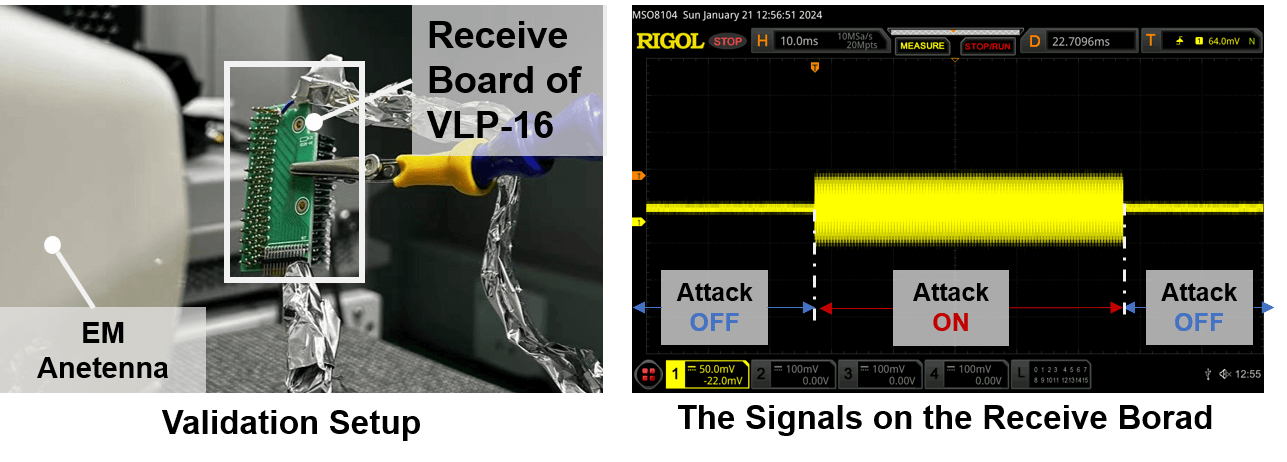}
        \vspace{-0.1in}

        \vspace{-10pt}
         \label{fig:Point_Interference_Validation}
     
	}
        \vspace{-0.1in}
	\caption{Points Interference.}
        \label{fig:Point_Interference}
	\vspace{-0.1in}
        \belowdisplayskip=12pt
\end{figure}

\textbf{Principle Analysis for Point Interference:} We propose that the \textit{direct attack} is the primary principle of point interference. As detailed in Sec.~\ref{sec:lidarsystembackground}, a benign LiDAR signal is typically a laser pulse.  When the LiDAR emits a laser pulse, its time-of-shooting and direction are registered. The laser pulse travels through the air until it hits an obstacle which reflects some of the energy, which is returned laser pulse. The distance of the object is calculated by measuring the time interval, so-called time of flight, between the emitted and the returned laser pulse. The time of acquisition is acquired according to the peak of the laser pulse. The point interference principle is illustrated in Fig.~\ref{fig:point_interference_principle}, the EMI can introduce noise into the benign signal by coupling the interfering signal into the wires between the transducer and amplifier. The interfering signal, when superimposed on the benign signal, alters the peak of the echo signal, thereby affecting the LiDAR's ranging.
Besides, as shown in Fig.~\ref{fig:points_interference_pattern}, we have also observed that when we slightly change the frequency, the interfering patterns will change significantly. This phenomenon is because the injected EM signal approaches or exceeds 1~GHz, and the sample rate of VLP-16's ADC is only 500~MHz, undersampling can lead to aliasing. This results in minor frequency changes altering the interference pattern, a phenomenon that has also been discovered in a concurrent prior work~\cite{bhupathiraju2023emi}. 
In our study, owing to the use of different frequencies and powerful devices, we can induce a distance error of up to 10~cm as illustrated in Fig.~\ref{fig:points_interference_pattern}, compared to the 4~cm error reported in the prior work~\cite{bhupathiraju2023emi}.

\textbf{Principle Validation for Point Interference:} To validate that the signal could couple into the LiDAR's receiver, we disassembled the LiDAR and extracted its receiving module. We then emitted a 990MHz EM signal towards the receiving module, a frequency capable of causing point interference. We observed that electromagnetic interference was indeed coupled into the transmission line of the receiving board as shown in Fig.~\ref{fig:Point_Interference_Validation}.

\subsubsection{Why Points Removal can be Induced}

\label{sec:analysis_Point eliminate and LiDAR poweroff}
\

\textbf{Principle Analysis for \textit{Points Removal}:} We propose \textit{direct attack} and \textit{indirect attack} are both potential principles to induce \textit{Points Removal}. 
\begin{itemize}
    \item Attack principle 1 (\textit{direct attack}): if we can inject high-amplitude EM signal into receiving circuit, it may saturate the receiving circuit and make the real laser pulse undetectable. 
    \item Attack principle 2 (\textit{indirect attack}): if we can compromise temperature sensor or hall effect sensor, it may induce LiDAR to detect L1 fault, leading LiDAR to consider some or all of the points as invalid. 
\end{itemize}




    

\begin{figure}[tp]
	\centering
	\subfigure[\textbf{The diagnostic interface of VLP-16 LiDAR.} Under 1.25~GHz EMI, the value of temperature and voltage will will significantly deviate from their normal values.]{
        \includegraphics[width=0.95\linewidth]{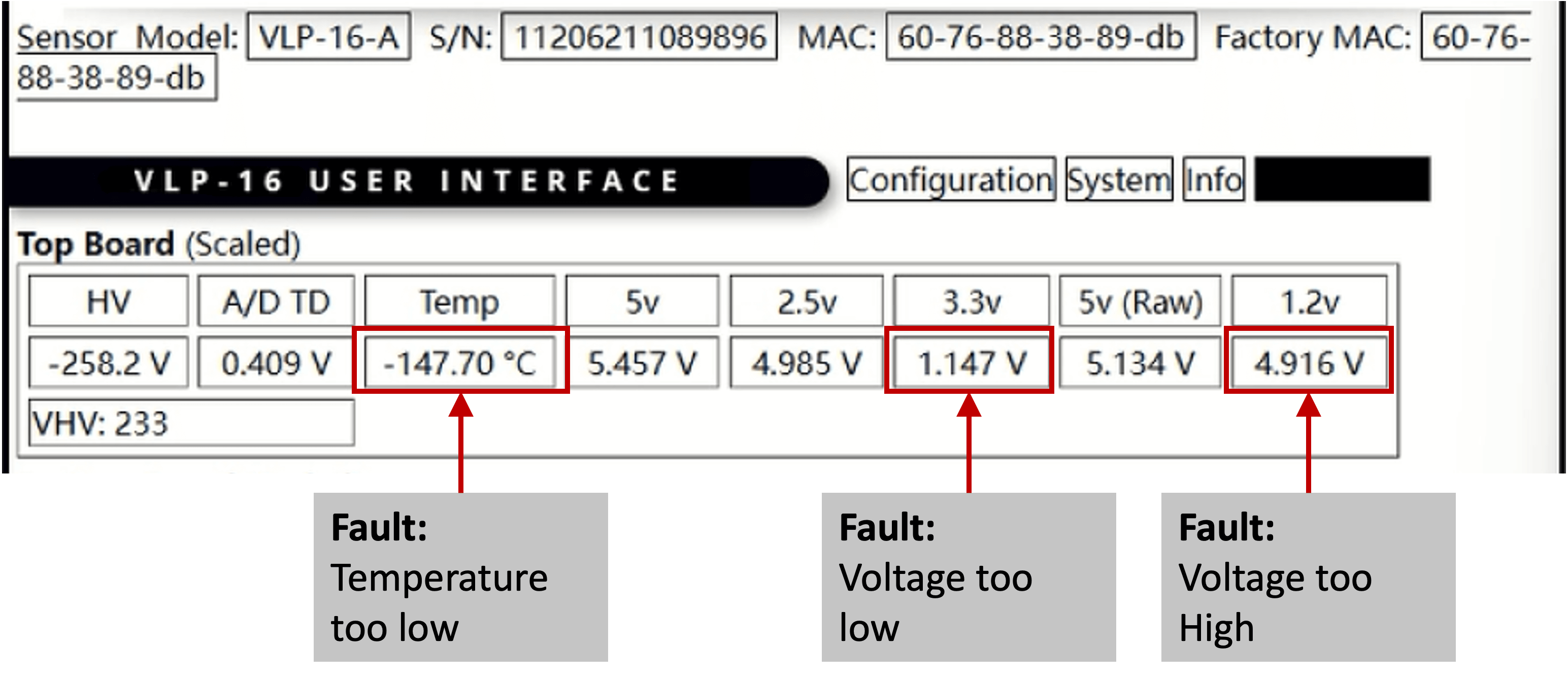}
         \vspace{-0.1in}

        \vspace{-10pt}
         \label{fig:velodyne_diagnostic}
	}
 
	\subfigure[\textbf{The output of temperature sensor}.When EMI is off, the temperature is around 40 $^{\circ}$C. When EMI is on, the temperature is fluctuate between -200$^{\circ}$C and 150$^{\circ}$C.]{
	\includegraphics[width=0.95\linewidth]{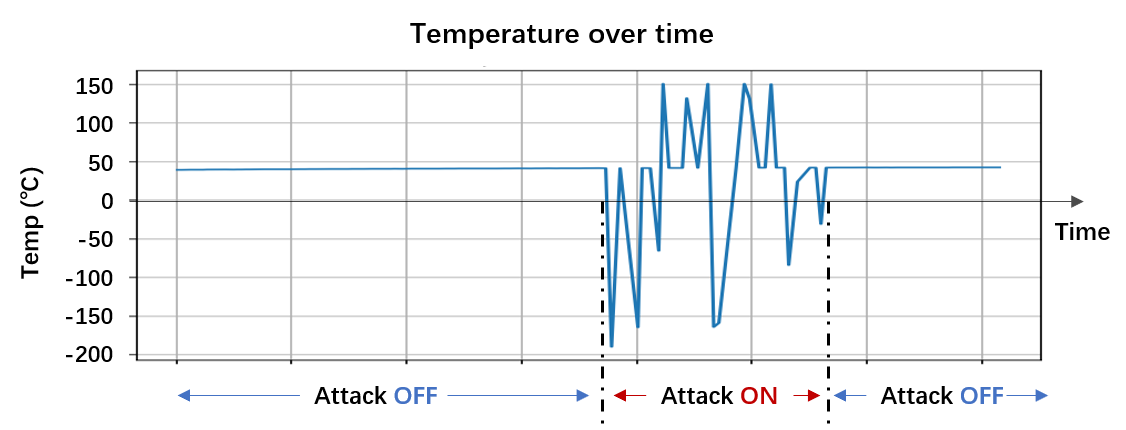}

         \label{fig:temperature_over_time}
     
	}
        \caption{Points Removal}
        \label{fig:velodyne_point_eliminate_diagnostic}
	\vspace{-0.1in}
\end{figure}

\textbf{Principle Validation for \textit{Points Removal}:} 
We validate the two principles through the Hypothetico-deductive method. If Principle 1 (direct attack) is correct, then when observing \textit{Points Removal} at a certain signal strength and gradually reducing the signal strength, we should observe a gradual recovery of the removed point cloud due to the decreasing amplitude of interference signals coupled to the receiving circuit.   If Principle 2 is correct, then when gradually reducing the signal strength to a certain value, a sudden recovery of the point cloud will be observed because at this point, the FDD mechanism no longer detects errors. Experimental evidence demonstrates that at certain frequencies, such as 1.1 GHz, we can indeed observe the gradual recovery of the removed point cloud, confirming that Principle 1 is correct at this frequency. However, at certain frequencies, such as 1.2 GHz, as the EM amplitude decreases, we observe a sudden recovery of the point cloud, indicating that at this frequency, Principle 2 is correct. 

In addition to the Hypothetico-deductive method, we further validate the Principle 2 by directly observing the readings of the monitoring sensor through the fault diagnostic interface.
As shown in Fig.~\ref{fig:velodyne_diagnostic}, the diagnostic information can be viewed through the fault diagnostic interface of VLP-16. We find that when EMI causes the points removed, some readings in the diagnostic interface exhibit anomalies. First, as shown in Fig.~\ref{fig:temperature_over_time}, the temperature sensor readings, which should be around 40 Celsius degree, fluctuate between -200 and 150 degrees. Second, some voltage rail readings deviate from normal values. We highly recommend readers to watch the demo video of fault diagnostic when \textit{Points Removal} occurs on the website~\cite{PhantomLiDAR}.


\begin{figure}[tp]
	\centering
        \subfigure[\textbf{The Rotational Speed of LiDAR before Powering Off.} When conduct LiDAR Power-off attack, the rotational speed of VLP-16 LiDAR significantly decreases, then leading to a denial of service, and ultimately resulting in powering off.]{
        \includegraphics[width=0.95\linewidth]{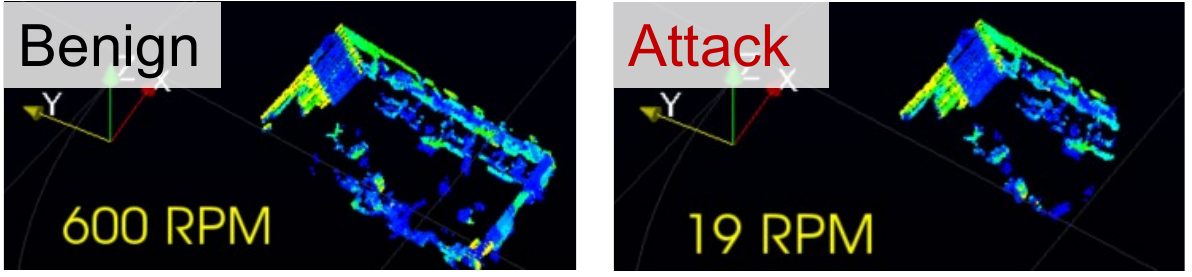}
         \vspace{-0.1in}

        \vspace{-10pt}
         \label{fig:RPM}
	}
 

        
	\subfigure[\textbf{Principle Validation.} The output of optical encoder AEDR-850X can be manipulated by EM signal.]{
	\includegraphics[width=0.95\linewidth]{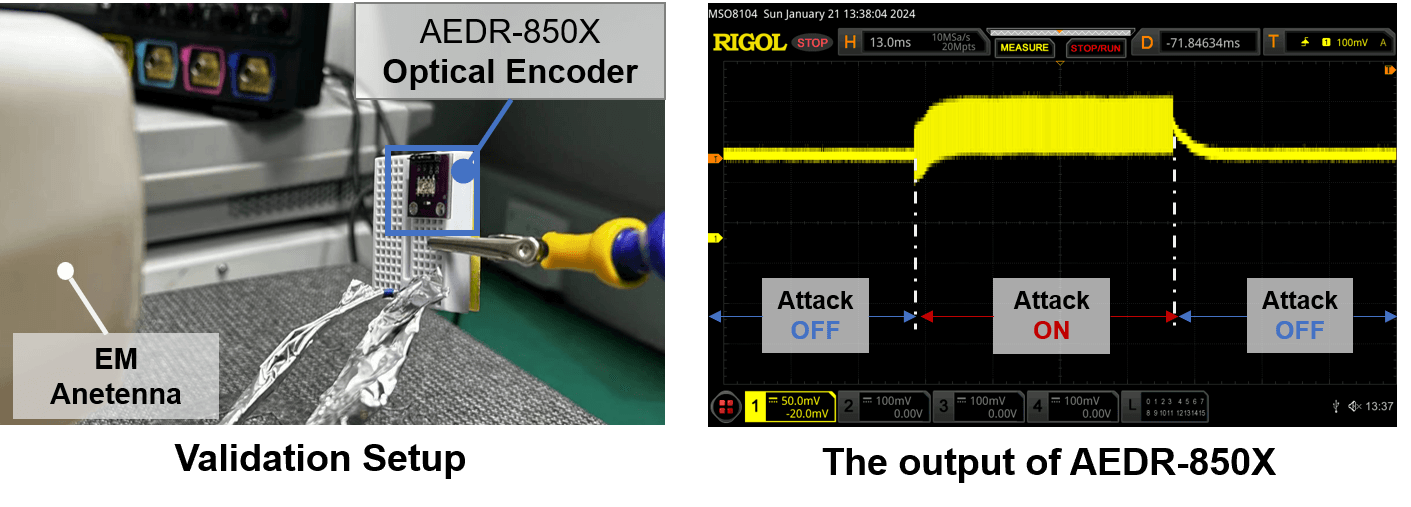}
        \vspace{-0.1in}

        \vspace{-10pt}
         \label{fig:Optical_Encoder_test}
	}
        \vspace{-0.1in}
	\caption{LiDAR Power-off.}
        \label{fig:}
	\vspace{-0.1in}
        \belowdisplayskip=12pt
\end{figure}

\begin{figure*}[pt]
    \centering
    \includegraphics[width=1\textwidth]{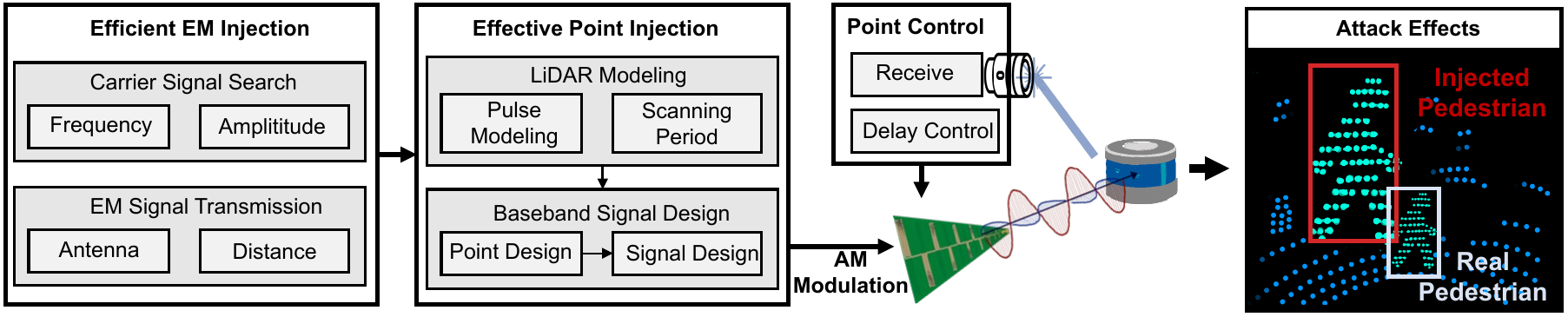}
    \vspace{-0.1in}
    \caption{The Work Flow of Controllable Point Injection. \textmd{First, the adversary identifies the optimal conditions for carrier signal and transmission which, when combined, can achieve the maximum intensity of EM injection. Subsequently, the adversary models the LiDAR parameters and designs a baseband signal capable of injecting points. After synchronization based on the receiver and delay control is completed, the baseband signal is modulated onto the carrier signal through amplitude modulation to generate the attack signal. This attack signal is capable of injecting controllable points.}}
    \vspace{-10pt}
     \label{fig:attack_design}
    
\end{figure*}

\subsubsection{Why LiDAR Power-off can be Induced}
\

\textbf{Principle Analysis for \textit{LiDAR Power-off}:} 
When EMI causes the LiDAR to power off, and we subsequently stop the attack and reboot the LiDAR, it still operates normally. Therefore, it is clear that we are not causing the LiDAR to power off by damaging the hardware. From Sec.~\ref{sec:lidarFDDbackground} and Fig.~\ref{fig:fault_diagnostic_management}, we learned that  the LiDAR may be forced to shut down when it encounters recoverable severe errors. Therefore, we propose that the \textit{indirect attack} is the principle of \textit{LiDAR Power-off}. Specifically, the \textit{LiDAR Power-off} is triggered by interfering the LiDAR's beam steering module.


\textbf{Principle Validation for \textit{LiDAR Power-off}:} 
We utilize Wireshark to record communication data of VLP-16 from the onset of the EM attack until the LiDAR shuts down. We find that in addition to some faults that also occurred during \textit{Points Removal}, the LiDAR's rotation speed exhibited severe anomalies. As shown in Fig.~\ref{fig:RPM}, despite being preset to 600 revolutions per minute (RPM), the speed will drop to 19 RPM, a reduction of 96.7\%. Based on our analysis above, such a drastic reduction in speed constitutes a severe fault, highly likely to cause the LiDAR to report an L2 fault and enter the power-off state. We further investigate which part of the beam steering module was disrupted by the EM signal. As shown in Fig.~\ref{fig:Optical_Encoder_Teardown}, upon disassembling the LiDAR, we discovered that the VLP-16’s beam steering module is a rotating motor monitored by an optical encoder AEDR-850X. As shown in Fig.~\ref{fig:Optical_Encoder_test}, we attacks an AEDR-850X with sinusoidal EM signals of 1050~MHz, which is the EM frequency inducing LiDAR Power-off. The test result reveals that EM signals can interfere with the data perceived by the optical encoder, which probably leads the LiDAR to erroneously detect an abnormally low motor speed, thereby causing a power-off.










\section{Controllable Point Injection Design}
\label{sec:points_injection}

In this section, we focus on how to inject and precisely manipulate fake (legitimate but erroneous) points with EMI against LiDAR systems. To execute a controllable point injection attack, it is crucial to address the following questions:
\begin{itemize}
\item  Q1. How to effectively inject EM signal? 

\item  Q2. How to inject fake points?

\item  Q3. How to control fake points?

\end{itemize}

For Q1, based on the preliminary study in Sec.~\ref{sec:feasibility study},  the most direct method for precise point cloud manipulation is to inject EM signals into the Time-of-Flight (ToF) circuit, rather than through fault management mechanisms to affect the point cloud. Therefore, the principle of \textit{Points Interference} attacks can be utilized, which is manipulating echo signal of LiDAR to control points.  In addition, we investigate the factors that influencing the effectiveness and efficiency of EM injection, including the frequency and amplititude of carrier signals and the types of antenna in different attack distances. Our goal is to identify the optimal conditions for these factors which, when combined, can achieve the maximum intensity of EM injection.

For Q2, the fake points can be injected by forging laser signal echoes. However, unlike the direct forgery of signals in laser attacks~\cite{jin2022pla,cao2019adversarial,shin2017illusion}, the transmission and coupling of pulse-shaped electromagnetic waves encounter significant attenuation and distortion. Therefore, we employ signal amplitude modulation (AM), using the designed laser pulse signal as the baseband signal, modulated onto a sinusoidal carrier signal of a resonant frequency.

For Q3, we reference methods from laser-based attack~\cite{jin2022pla}, using photoelectric sensors to detect the operational status of LiDAR systems, then set a precise delay to control the timing of EM signal emission, ultimately achieving control over the point cloud.

The workflow of controllable points injection is shown in Fig.~\ref{fig:attack_design}. The \textbf{Efficient EM Injection} module identify the optimal conditions for carrier signal and transmission which, when combined, can achieve the maximum intensity of EM injection. The \textbf{Effective Point Injection} Module models the LiDAR parameters and design baseband signal that can inject points. The \textbf{Point Control} module introduces a closed-loop feedback mechanism integrating receive and delay control.

\subsection{Efficient EM Injection}
We utilize a sinusoidal EM signal as the carrier wave. The stronger the intensity of the EM injection, the more likely the forged echo signal is to be perceived as a valid echo, thereby increasing the success rate of fake points injection. Consequently, we are motivated to study the main factors that affect the effectiveness and efficiency of EM injection.

\subsubsection{Carrier Signal Search} The coupling efficiency between a victim circuit and an EM signal is determined by the frequency and amplitude of the EM signal. 

\textbf{Frequency.} When conducting IEMI attacks, we consider the wires in the victim LiDAR system as receiving antennas. The effective EM signal frequency can be estimated based on the electrical length of the targeted/selected antenna~\cite{redoute2009emc,wang2022ghosttouch}. Empirically, if the length of the targeted antenna is $l$, the optimal coupling frequency of the EM signal lies between $\frac{c}{50l}$and $\frac{c}{2l}$, where $c$ is the light speed. The specific optimal coupling frequency also depends on the shape, material, and other characteristics of the targeted antenna. In realistic attack scenarios, attackers can utilize the aforementioned theory to establish an approximate frequency range and subsequently identify the resonant frequency with higher coupling efficiency through frequency sweep.

\textbf{Amplititude.} The amplitude is a critical factor in EMI, it is generally considered that higher signal amplitude results in stronger interference. In practical experiments, there may be instances where increasing the device's displayed amplitude does not enhance interference. For example, in our experiments, the signal generator's maximum amplitude is 19~dBm (80~mW), and the amplifier is 50~W. However, we observed that changing the signal generator's amplitude (e.g., from 10~dBm to 19~dBm) sometimes does not affect the final output. This is due to the saturation characteristics of the amplifier, which limits the maximum output strength of the signal.

\subsubsection{EM Signal Transmission}
Firstly, the frequency range of the antenna must cover the range of the sweep. Secondly, it is advantageous for the antenna to have as high a gain as possible to reduce signal attenuation. In our experiments, to accommodate various attack distances, we selected two types of transmission antennas. For attack distances less than 10~cm, we choose a near-field probe to generate the EM signal due to its greater portability. For attack distances greater than 10 cm, we opt for a log-periodic antenna due to its superior directionality.

\subsection{Effective Point Injection}
We inject fake points by forging LiDAR signal echoes. Firstly, we model the LiDAR system, then design the pulsed signal based on the points we intend to inject. Finally, we modulate the pulsed signal onto a sinusoidal carrier signal.

\subsubsection{LiDAR Modeling} 
For the target victim LiDAR, we need to acquire the necessary information to guide signal design. First, we must obtain its signal waveform, typically consisting of one or multiple pulses. For instance, the signal of VLP-16 is a pulse with a width of about 10 $ns$. Then, we need to determine the LiDAR's scanning cycle, which is usually available in the data manual. For example, the VLP-16 operates on a cycle of 55.296 $\mu s$, during which 16 laser lines are emitted and received at intervals of 2.304 $\mu s$ in a specific sequence, followed by a recharge period of 18.432 $\mu s$.

\subsubsection{Baseband Signal Design} 
The design process of the attack signal involves converting the intended point cloud into a series of pulses, where each pulse corresponds to a spoofing point, and the timing of each pulse's peak defines the spatial coordinates of the spoofing point. The method for designing the baseband signal is akin to that used in previous laser-based attacks~\cite{jin2022pla}. However, using the baseband signal directly is ineffective for injecting the signal into LiDAR systems through EMI. Therefore, we use the attack signal as the baseband signal and modulate it onto a carrier signal using amplitude modulation (AM) to enable effective injection.

\subsection{Point Control}

With the AM-modulated EM pulses, we can inject spoofing points into the LiDAR. However, However, these spoofing points are disorganized and cannot be stabilized in a fixed pattern and position, as shown in Fig.~\ref{fig:synchronization} in Appendix. To achieve controllable injection, it is necessary to synchronize the signal with the LiDAR's scanning sequence. Inspired by previous laser-based attack~\cite{jin2022pla}, We introduce an integrated receive-delay-fire closed-loop feedback mechanism for point cloud control. we utilize a photoelectric sensor to detect the operational cycle of the LiDAR and adopt the delay control function in signal generator to achieve synchronization.





\section{Evaluation}
\label{evaluation}

In this section, We evaluate the effectiveness of the PhantomLiDAR in terms of four types of attacks: \textit{Points Interference, Points Removal, LiDAR Power-off} and \textit{Points Injection}.

\subsection{Evaluation Overview}

\subsubsection{Methodology}
As each attack possesses distinct characteristics, we designed unique evaluation methods for each attack to better understand their strength and limitations.

Firstly, in Sec.~\ref{sec:attack_different_LiDAR}, we focus on the \textit{Points Interference, Points Removal,} and \textit{LiDAR Power-off} attacks,  as these three attacks posing greater potential threats in real-world scenarios due to their simple implementation. We evaluated the effectiveness of these attacks on five LiDARs and discovered that different LiDARs have varying vulnerable frequencies and show different robustness to IEMI. This experiment may provide insights for future LiDAR design or contribute to the establishment of new electromagnetic compatibility (EMC) testing standards.

In Sec.~\ref{sec:evaluation_point_interference}, we evaluate the \textit{Points Interference} attack on five LiDAR-based 3D object detection models. We synthesize the \textit{Points Interference} datasets based on the KITTI~\cite{KittiBenchmark} dataset, which is a large-scale dataset collected from the real world. By conducting experiments on two LiDAR-based models and three fusion-based models, we observed that point cloud interference attacks lead to a degradation in the performance of 3D object detection models.

In Sec.~\ref{sec:evaluation_point_eliminate}, we evaluate the \textit{Points Removal} attack in the real world.  As Point Removal can completely erase point clouds, it enables the \textit{Hiding} attacks on object detection models, making specified target objects undetectable. With the attack goal of \textit{Hiding} attack, we focus on evaluating the attack impacts of the attacker's distance and angle. Additionally, we evaluated the aiming requirements for EMI attacks. This evaluation highlights an advantage of EM attacks over laser attacks with the reduced need for precise aiming.

In Sec.~\ref{sec:evaluation_LiDAR_poweroff}, for the \textit{LiDAR Power-off} attack, we evaluate the impact of attack distance and discuss its potential threats in realistic scenarios.

In Sec.~\ref{sec:evaluation_point_injection}, we evaluate \textit{Point Injection} from the maximum number of injected points and the precise control over the points, thereby verifying the effectiveness of the attack.

In Sec.~\ref{sec:evaluation_moving}, we explore the feasibility of hiding a targeted object when the victim LiDAR is in motion. 

\subsubsection{Attack Devices}
All experiments in this section utilized shared attack attack devices, including a Keysight N5712b vector signal generator (32K USD)~\cite{n5172b} for EMI signal generation, a Mini-Circuits HPA-50W-63+ power amplifier(20K USD)~\cite{hpa50w} for EMI signal amplification, and a log-periodic antenna for remote signal transmission. The log-periodic antenna, valued at approximately 15 USD~\cite{Log_Antenna}, has a frequency range of 600MHz to 6000MHz and a gain of 15 dBi.
Additional devices specific to \textit{Points Injection} attacks are detailed in Sec.~\ref{sec:evaluation_point_injection}.

\subsection{Attack on Different LiDARs}
\label{sec:attack_different_LiDAR}

\subsubsection{Victim LiDAR:} We evaluate EMI attacks on five off-the-shelf LiDARs as shown in Fig.~\ref{fig:LiDARs}, which include three mechanical LiDARs (VLP-16~\cite{velodyneVLP16}, RS-16~\cite{robosenseRS16}, RS-Bpearl~\cite{robosenseRSBpearl}) and two MEMS LiDARs(RS-M1~\cite{robosenseM1},RS-M1P~\cite{robosenseM1P}). The RS-M1 and RS-M1P, serve as primary LiDARs for perception, and are widely deployed in both currently released~\cite{ZEEKER007,YangwangU8,XIAOPENGG9} and upcoming~\cite{RS-M1MoreThan60Models} vehicles.

\subsubsection{Attack Setup} We put the antenna 30~cm away from LiDARs, and conduct a frequency sweep test on these LiDARs from 500~MHz to 3500~MHz with a step of 5MHz. The output amplitude of the signal generator is 0~dBm and the amplifier is 50~W. 

\begin{table}[]
\renewcommand\arraystretch{1.2}
 \centering
  \caption{Attacks on Five LiDARs}
  \label{tab:attack_result}
  \vspace{-0.1in}
  \begin{threeparttable}
    \resizebox{\linewidth}{!}{
    \begin{tabular}{c|c|c|c|ccc}
    \hline
    \toprule
    \multirow{2}{*}{\textbf{LiDAR}} & \multirow{2}{*}{\textbf{\begin{tabular}[c]{@{}c@{}}Beam \\ Steering\end{tabular}}} & \multirow{2}{*}{\textbf{\begin{tabular}[c]{@{}c@{}}Laser\\ Pulse\end{tabular}}} & \multicolumn{1}{l|}{\multirow{2}{*}{\textbf{\begin{tabular}[c]{@{}l@{}}Timing\\ Random\end{tabular}}}} & \multicolumn{3}{c}{\textbf{Attack}}                                                                                                                                                                                \\ \cline{5-7} 
                                    &                                                                                    &                                                                                 & \multicolumn{1}{l|}{}                                                                                  & \textbf{\begin{tabular}[c]{@{}c@{}}Points \\ Interference\end{tabular}} & \textbf{\begin{tabular}[c]{@{}c@{}}Points \\ Removal\end{tabular}} & \textbf{\begin{tabular}[c]{@{}c@{}}LiDAR \\ Power-off\end{tabular}} \\ \hline
    VLP-16                          & Rotary                                                                             & Single                                                                          & N/A                                                                                                      & \checkmark                                                                      & \checkmark                                                                   & \checkmark                                                                   \\
    RS-16                           & Rotary                                                                             & Single                                                                          & Yes                                                                                                      & \checkmark                                                                      & \checkmark                                                                   & --                                                                   \\
    RS-Bpearl                       & Rotary                                                                             & Encode                                                                          & Yes                                                                                                      & --                                                                      & \checkmark                                                                   & --                                                                   \\
    RS-M1                           & MEMS                                                                               & Encode                                                                          & Yes                                                                                                      & \checkmark                                                                      & \checkmark                                                                   & \checkmark                                                                   \\
    RS-M1P                          & MEMS                                                                               & Encode                                                                          & Yes                                                                                                      & \checkmark                                                                      & \checkmark                                                                   & --                                                                   \\ 
    \bottomrule \hline

    \end{tabular}}
    
    \begin{tablenotes}
      \item {\small \checkmark ~ Attack Succeed  \hspace{0.02in}   -- ~ Not Observed \hspace{-0.02in}}
    \end{tablenotes}

 \end{threeparttable}
\end{table}

\begin{figure}[pt]
    \centering

    \includegraphics[width=\linewidth]{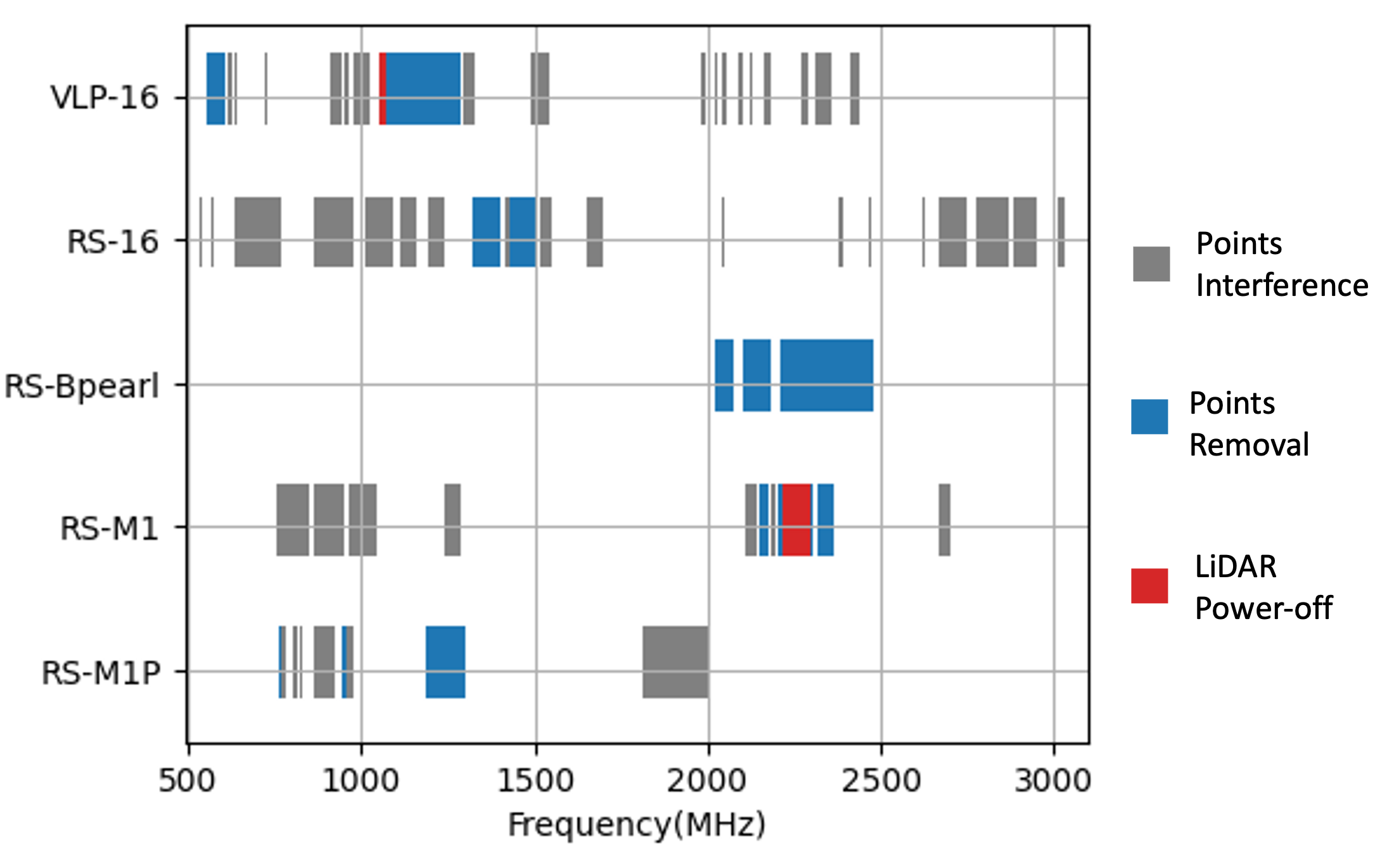}
    \vspace{-0.2in}
    \caption{Attack effects on five LiDARs and the frequency corresponding to each attack}
     \label{fig:QualitativeSweepFrequency}
    
\end{figure}

\subsubsection{Attack Results} The attack results are shown in Table~\ref{tab:attack_result}. The relationship between attack effects and EMI frequency is shown in Fig.~\ref{fig:QualitativeSweepFrequency}.  It is observed that LiDARs with different structures exhibit varying vulnerabilities and susceptible frequencies. We did not observe the phenomenon of \textit{Point Interference} attacks on the RS-Bpearl, indicating that its EM protection for received signals is superior compared to the other LiDAR models. \textit{Points Removal} was observed in all LiDAR, but it is important to note that on the VLP-16 and RS-16, EMI could erase all point clouds. However, on the other three LiDAR models, we only observed partial removal of point clouds, which we speculate is due to the directionality of EM waves affecting only certain circuits. As for the \textit{LiDAR Power-off} attack, it was observed only on the older models, VLP-16 and RS-M1, suggesting that newer generations of LiDAR have optimized diagnostic and management of serious faults.

\subsection{Points Interference}
\label{sec:evaluation_point_interference}
\subsubsection{Impact of Signal Amplitude}

To systematically explore the influence of amplitude on point interference, we conducted further experiments on the VLP-16 LiDAR. First, we selected two typical frequencies: 989MHz and 990 MHz. As shown in Fig.~\ref{fig:points_interference_pattern}, 989 MHz can produce a sinusoidal pattern, whereas 990 MHz can generate a random pattern. We then increased the signal generator's amplitude from -40 dBm to 0 dBm and connected it to an amplifier with a gain of 56 dB and a maximum output of 50 W (47 dBm). By recording the distance error at different amplitudes, we obtained the results shown in Fig.~\ref{fig:amp_vs_distanceerror}. It can be seen that as the amplitude increases, the distance error gradually increases and the trend is consistent across different frequencies. When the signal generator's amplitude reaches approximately -10 dBm, the distance error reaches around 14 cm at most. However, despite further increases in the signal generator's output, the distance error did not increase significantly. This is primarily because the amplifier output reached saturation, preventing accurate control of the final signal amplitude by the signal generator. Furthermore, the saturation distortion caused by the amplifier resulted in signal distortion after -10 dBm, which explains the decrease in distance error around -10 dBm.

\subsubsection{Evaluation on Large-scale Dataset}

To systematically evaluate the impact of Point Interference on 3D object detection models, we simulated the attack based on the large-scale dataset and then input the corrupted dataset into the detection models.

\begin{figure}[pt]
    \centering

    \includegraphics[width=0.95\linewidth]{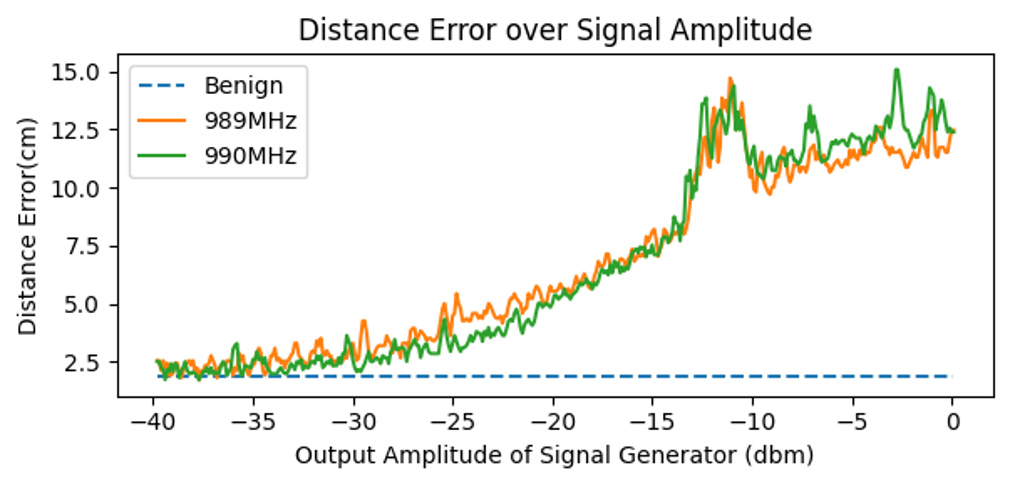}
    \vspace{-0.1in}
    \caption{Impacts of signal amplitude to distance error of \textit{Points Interference}. \textmd{As the amplitude increases, the distance error gradually increases, and the trend is consistent across different frequencies.}}
    \label{fig:amp_vs_distanceerror}
    
\end{figure}

\textbf{Dataset:} We synthesize the attack on KITTI dataset, which comprises diverse real-world driving scenarios. Based on Sec.~\ref{sec:Point_Interference_analysis}, the Point Interference can be formulated as the following equation:

\begin{equation}
    \begin{cases} 
    & r_{i}' = r_{i} + Random[-\varepsilon,\varepsilon ]\\
    & (r_{i} ,\theta_{i}  ,\varphi_{i}  )\in \mathbb{PC}, i\in[1,n]\\
    & (r_{i}' ,\theta_{i}  ,\varphi_{i}  )\in \mathbb{PC'}, i\in[1,n],
    \end{cases}
\end{equation}
where $\mathbb{PC}$ stands for benign point cloud and $\mathbb{PC'}$ stands for compromised point cloud. We represent a point with spherical coordinates $(r ,\theta ,\varphi )$, where $r$ denotes the radial distance, $\theta$ the polar angle, and $\varphi$ the azimuthal angle. EM interference affects only $r$, introducing random noise within the range of $[-\varepsilon,\varepsilon ]$. As shown in Fig.~\ref{fig:amp_vs_distanceerror}, up to 10 cm distance error can be stably induced based on our attack device. Therefore, to ensure the physical realizability of the corrupted dataset, we synthesize the datasets when $\varepsilon$ is set at 5 cm and 10 cm, respectively.

\begin{figure}[pt]
    \centering

    \includegraphics[width=0.48\textwidth]{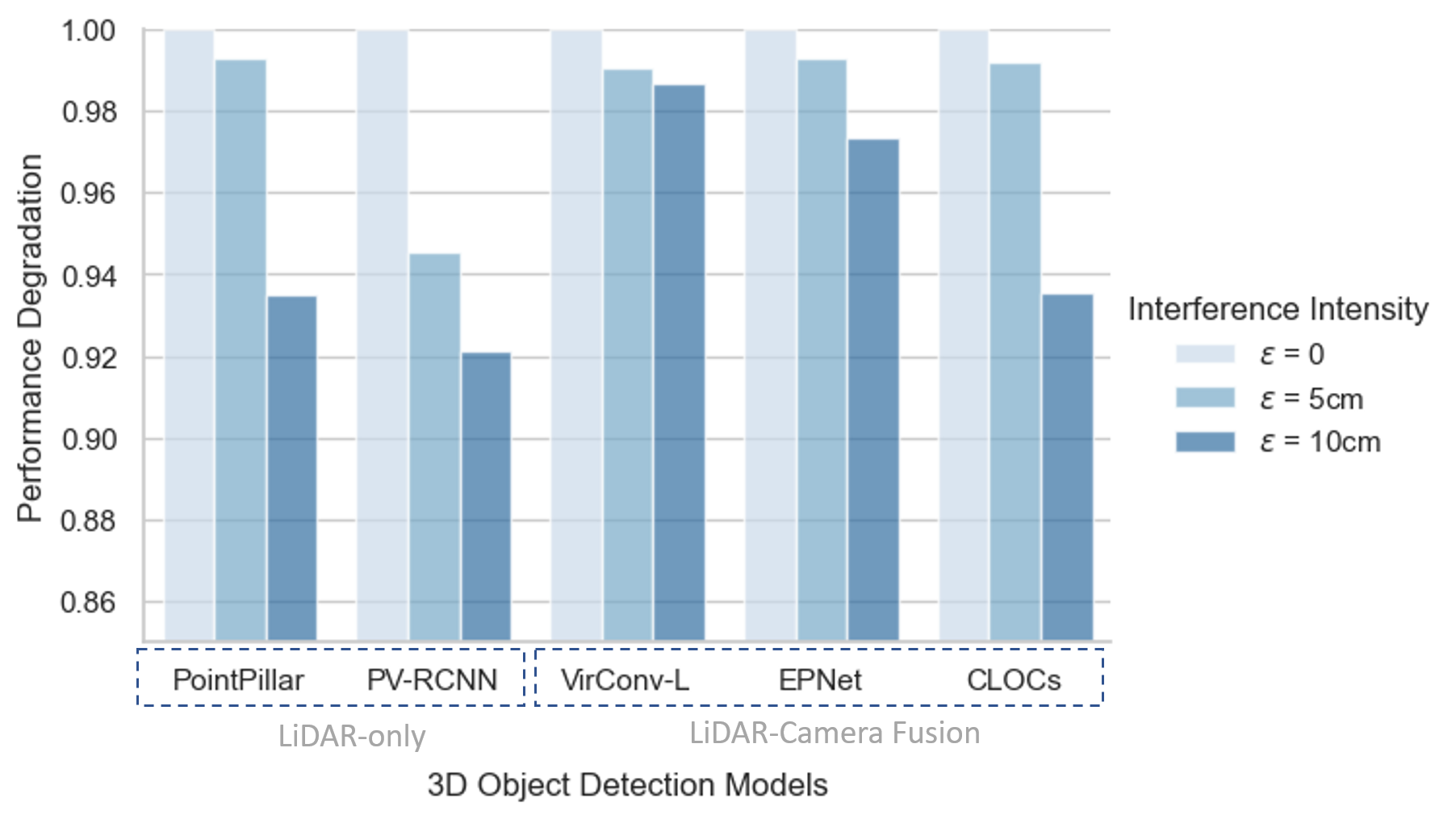}
    \vspace{-0.2in}
    \caption{Robustness (Rb) of five 3D objection detection models against Point Interference. \textmd{The robustness of fusion-based models is generally stronger than that of LiDAR-based models}}
    \vspace{-10pt}
     \label{fig:evaluation_point_interference}
    
\end{figure}

\textbf{Detection Models:} We evaluate the impact of \textit{Points Interference} on five 3D object detection models. Two are LiDAR-based models: PointPillar~\cite{lang2019pointpillars} and PV-RCNN~\cite{shi2020pv}. Additionally, there are three LiDAR-Camera fusion models: the early-fusion model VirConv-L~\cite{wu2023virtual}, the feature-fusion model EPNet~\cite{huang2020epnet} and the results-fusion model CLOCs~\cite{pang2020clocs}.

\textbf{Metrics:} 
Same as the KITTI benchmark~\cite{geiger2013vision}, we consider a car detection as successful when the IoU between a ground-truth 3D bounding box and a predicted 3D bounding box exceeds 0.7. Based on the IoU threshold, we use Average Precision(AP) to measure the overall detection performance of a model. AP is the official metric for the KITTI benchmark for a comprehensive assessment of a model's performance. We calculate the AP in the moderate difficulty, aligning with the KITTI benchmark~\cite{KittiBenchmark}, where models are ranked based on the moderately difficult results. 
To evaluate the model's robustness against interference, we introduce a new metric Robustness Coefficient $Rb$:
\begin{equation}
    Rb = \frac{AP'}{AP_{benign}}
\end{equation}
where $AP'$ and $AP_{begnin}$ denote the model's average precision on the interfered dataset and the benign dataset, respectively. 

\textbf{Results:} 
The APs of the five 3D Object detection models under different-intensity points interference are presented in Table.~\ref{tab:interfernce_result}. 
It is observed that with increasing interference intensity, the performance of the detection models decreases, demonstrating the attack devices with higher power can, demonstrating that devices with higher power have stronger attack effects.

The Rbs of models are shown in Fig.~\ref{fig:evaluation_point_interference}. The robustness of Fusion-based models is generally stronger than that of LiDAR-based models, indicating that sensor fusion has potential as a countermeasure against Points Interference attacks. Within the fusion-based models, CLOCs exhibits the least robustness. This is attributed to CLOCs being a result fusion approach that combines camera and LiDAR detection candidates before applying Non-Maximum Suppression (NMS). Such a loosely coupled fusion method is ineffective in defending against Points Interference attacks. Further, we note that even interference at 10cm does not significantly compromise the 3D object detection models(Rbs > 0.92), indicating that existing models have a certain level of robustness to interference. 
This experiment helps us objectively understand that the harmfulness of \textit{Points Interference}

\subsection{Points Removal}
\label{sec:evaluation_point_eliminate}

The attack goal of \textit{Points Removal} is to hide a target object against 3D object detection model. The efficiency of Point Removal in achieving this goal is unquestionable, as it can completely erase the points representing the target object. Therefore,  to gain a more comprehensive understanding of the attack's capabilities, we investigate the impacts of the attacker's location and aiming. 

\subsubsection{Experimental Setup}
\begin{figure}[pt]
    \centering

    \includegraphics[width=0.48\textwidth]{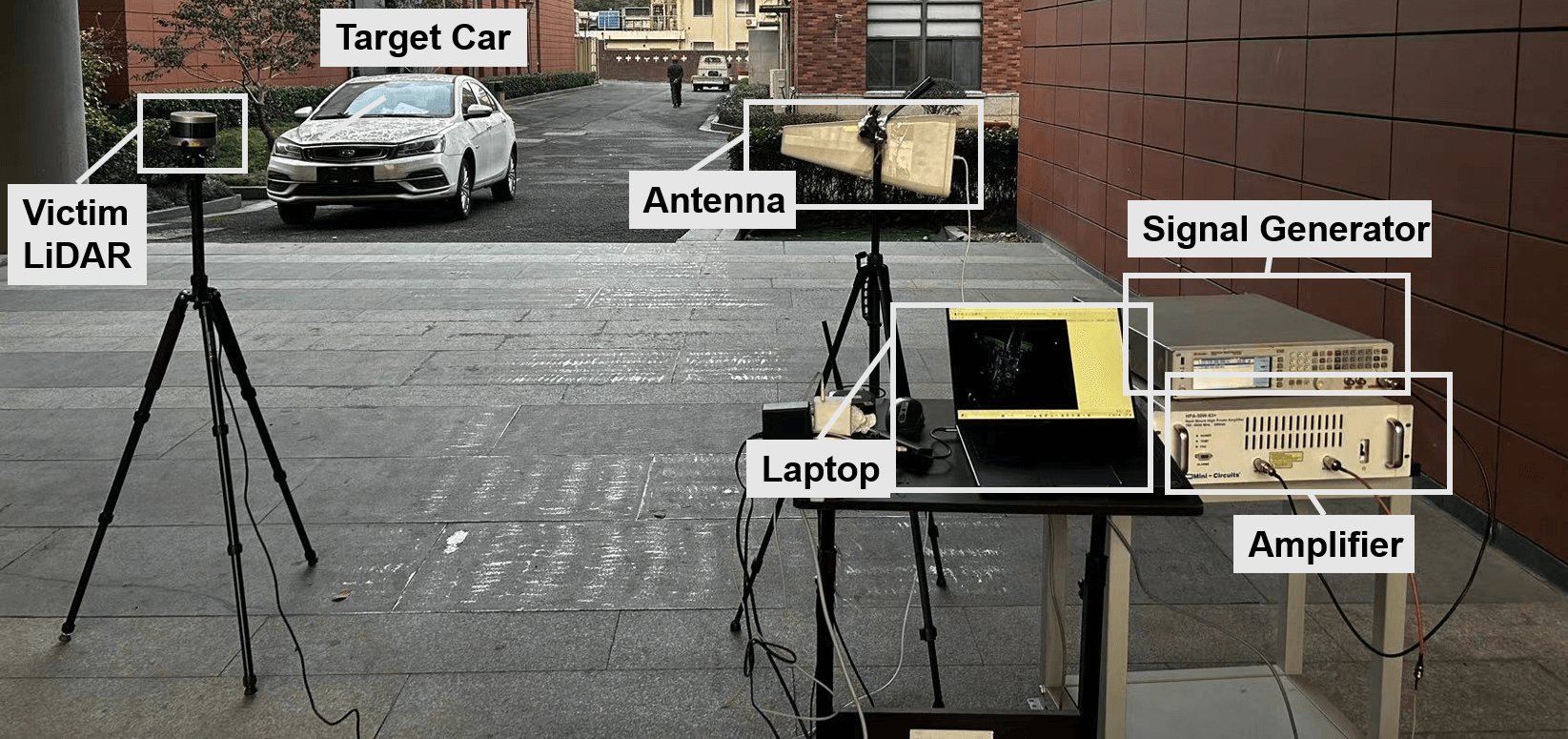}
    \vspace{-0.2in}
    \caption{The Attack Setup Outdoors. \textmd{The attack goal is to attack the victim LiDAR by IEMI, and make the target car undetectable on 3D object detection model.}}
    \vspace{-10pt}
     \label{fig:setup_outdoors}
    
\end{figure}

The experimental setup is shown in Fig.~\ref{fig:setup_outdoors}. We conducted physical experiments on campus roads. The attack target is a car, which is the most prevalent target in real autonomous driving scenarios. The victim LiDAR is VLP-16, and the detection model is PointPillars~\cite{lang2019pointpillars}.

\begin{figure}[tp]
	\centering
	\subfigure[Illustration of setup]{
        \includegraphics[width=0.4\linewidth]{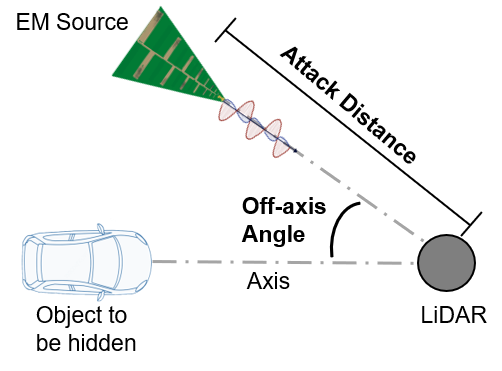}
         \label{fig:illustration_attacker_location}
	}
	\subfigure[Results]{
	\includegraphics[width=0.5\linewidth]{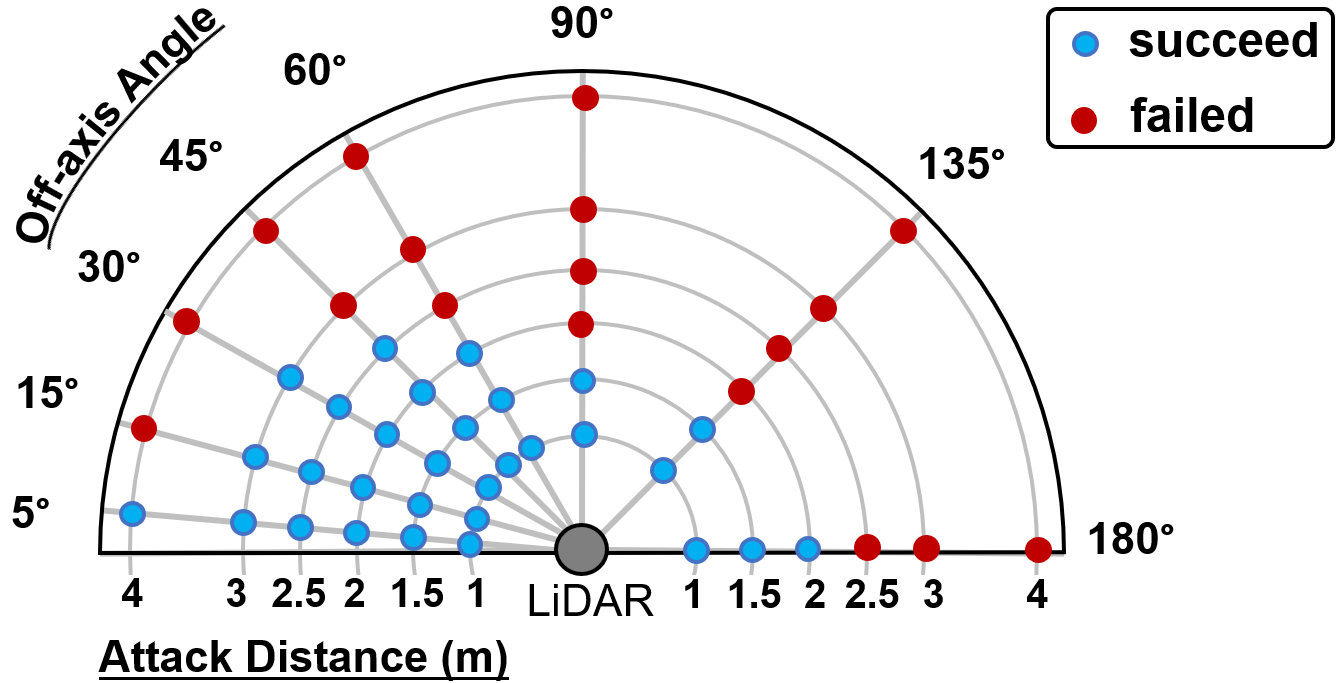}
         \label{fig:results_attacker_location}
	}
        \vspace{-0.1in}
	\caption{Impacts of attacker's location. \textmd{Within a distance of 1.5 meters, the attacker can hide the target object from any angle. When the off-axis angle is 5$^\circ$, the attack can succeed beyond 4 meters away (5.5 meters at most in this paper).}}
        \label{fig:impatcs_attacker_location}
	\vspace{-0.1in}
        \belowdisplayskip=12pt
\end{figure}

\subsubsection{Impacts of Attacker's Location}
As illustrated in Fig.~\ref{fig:illustration_attacker_location}, the attacker's location includes their off-axis angle and distance to the LiDAR and the target object.
A larger range of locations from which an attack can be successfully executed implies greater robustness and stealth of the attack in real-world scenarios. We investigated the attack’s effectiveness by conducting attacks on the LiDAR from distances of $[1, 1.5, 2, 2.5, 3, 4]$ meters at off-axis angles of $[5, 15, 30, 45, 60, 90, 145, 180]$ degrees. 

The results are shown in Fig.~\ref{fig:results_attacker_location}. It demonstrates that within a distance of 1.5 meters, the attacker can hide the target object from any angle, as all LiDAR point clouds are erased at this distance. Furthermore, we found that the smaller the off-axis angle, the greater the distance at which an attack can be successful. To find the longest attack distance, we move the EM source as far as possible.
With our test setup, the attack could be successfully executed when the EM source was up to 5.5 meters away from the LiDAR. We anticipate that the maximum effective attack distance could be more than 5.5 meters, particularly with the use of a high-power EM source.


    

\begin{figure}[tp]
	\centering
	\subfigure[Vertical Angle]{
        \includegraphics[width=0.45\linewidth]{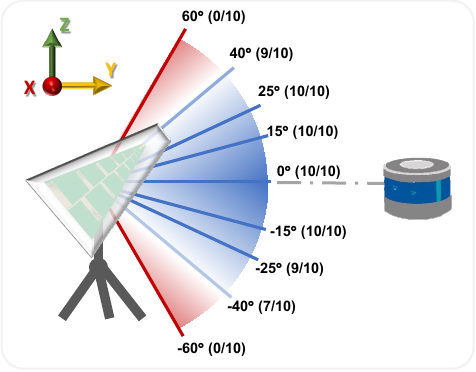}
         \label{fig:aiming_vertical}
	}
	\subfigure[Horizontal Angle]{
	\includegraphics[width=0.45\linewidth]{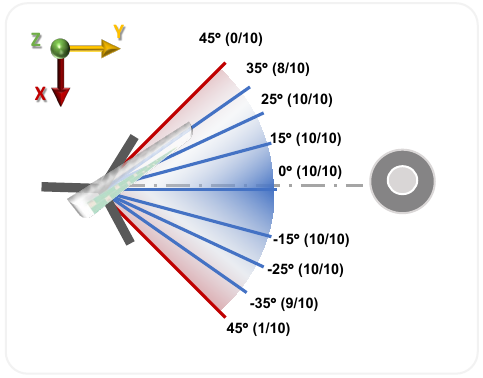}
         \label{fig:aiming_horizontal}
	}
        \vspace{-0.1in}
	\caption{Impacts of Aiming. \textmd{The EM antenna could deviate up to 40$^\circ$ vertically or 35$^\circ$ horizontally while still achieving a hiding attack effect. The results show very low aiming requirement for our attacks.}}
        \label{fig:aiming}
	\vspace{-0.1in}
        \belowdisplayskip=12pt
\end{figure}


    

\subsubsection{Impacts of Aiming}

Aiming is a common challenge faced in remote attacks in the real world, a problem that has been especially pronounced in previous laser attacks~\cite{jin2022pla}. In this experiment, we investigate the aiming requirements for EM attacks. The lower the aiming precision required for an attack, the more feasible it is in real-world scenarios. We kept the location of the EM antenna constant (1 meter from the LiDAR) and then varied the direction in which the EM antenna was pointed. The point clouds were then input into the detection model to observe whether the attack is successful. The results are shown in Fig.~\ref{fig:aiming}, the EM antenna could deviate up to 40° vertically or 35° horizontally while still achieving a hiding attack effect. This result suggests that EM attacks do not require precise aiming, an advantage over laser attacks.








\subsection{LiDAR Poweroff}
\label{sec:evaluation_LiDAR_poweroff}
As analyzed in Sec.~\ref{sec:analysis_Point eliminate and LiDAR poweroff}, the power-off is a protective mechanism employed by LiDAR systems when encountering serious faults. In real-world scenarios, remotely inducing a LiDAR power-off via IEMI could significantly impair the functionality of autonomous vehicles. 
To explore the realistic threats of \textit{LiDAR Power-off}, we evaluate the impacts of attack distance on VLP-16 and RS-M1. With the attack devices illustrated in Fig.~\ref{fig:setup_outdoors}, we were able to induce \textit{LiDAR Power-off} against VLP-16 and RS-M1 at distances of 30 cm and 50 cm, respectively. This distance is sufficient to pose a threat in real life, as it allows for attacks to be conducted from outside a vehicle.
We recommend readers to watch the demo video of \textit{LiDAR Power-off} available on the website~\cite{PhantomLiDAR}.

A conceivable attack scenario is that an attacker places the amplifier and signal generator inside their vehicle and extends the EM antenna outside through a long wire. When an autonomous vehicle stops, such as at a red light, the attacker could bring the EM antenna close to the LiDAR of the autonomous vehicle, causing it to power off. Then, even if the vehicle moves away, the power-off effect would persist.

\subsection{Points Injection}
\label{sec:evaluation_point_injection}
We evaluate \textit{Points Injection} from two perspectives: firstly, the maximum number of injected points, which has been an important metric in previous LiDAR attacks, and secondly, the control over the points, specifically whether we can inject a point cloud with a specified pattern.

\subsubsection{Setup}
 The setup of \textit{Point Injection} is shown in Fig.~\ref{fig:Point_Injection_Setup} in Appendix. In addition to a vector signal generator, an amplifier, an EM antenna, Point Injection also requires an arbitrary waveform generator to create the baseband signal and set the delay, and a photodetector to receive signals from the LiDAR. The LiDARs under test are VLP-16 and RS-16, which are the two LiDARs tested in SOTA laser-based attacks~\cite{jin2022pla}. 

\begin{figure}[pt]
    \centering

    \includegraphics[width=\linewidth]{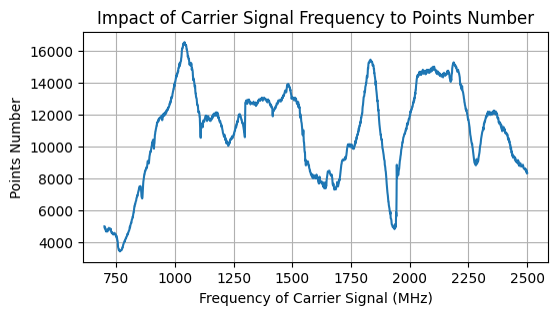}
    \vspace{-0.1in}
    \vspace{-10pt}
    \caption{Impact of Carrier Frequency to Points Number. \textmd{}}
    \vspace{-0.1in}
     \label{fig:freq_vs_pointsnumber}
    
\end{figure}

 \subsubsection{Number of Injected Points} 
The comparison of the number of injected points is only meaningful when the victim LiDAR models are the same and have same rotation speed. Therefore, we selected VLP-16 as the victim LiDAR, a LiDAR commonly used in previous point injection research~\cite{jin2022pla,cao2019adversarial,shin2017illusion}.

\begin{figure}[t]
	\centering
        \subfigure[Maximum Injected Points.]{
		\includegraphics[width=0.45\linewidth]{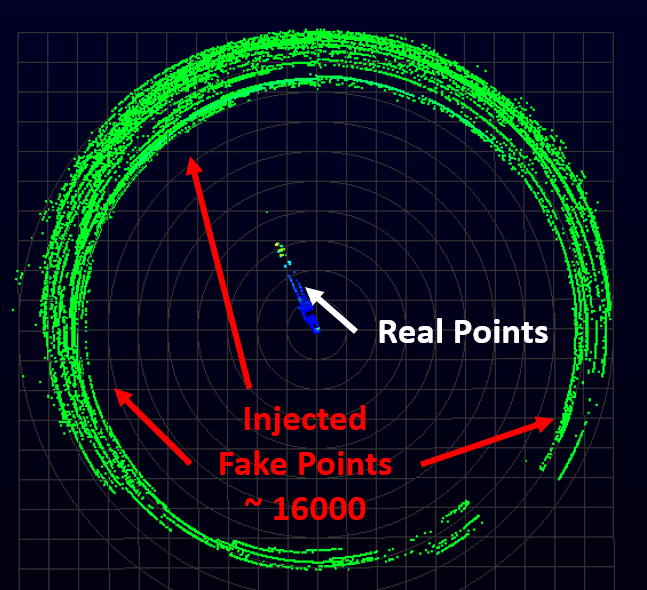}
		\label{fig:wall_injected}
	}	
	\subfigure[Specified Pattern Injection]{
		\includegraphics[width=0.45\linewidth]{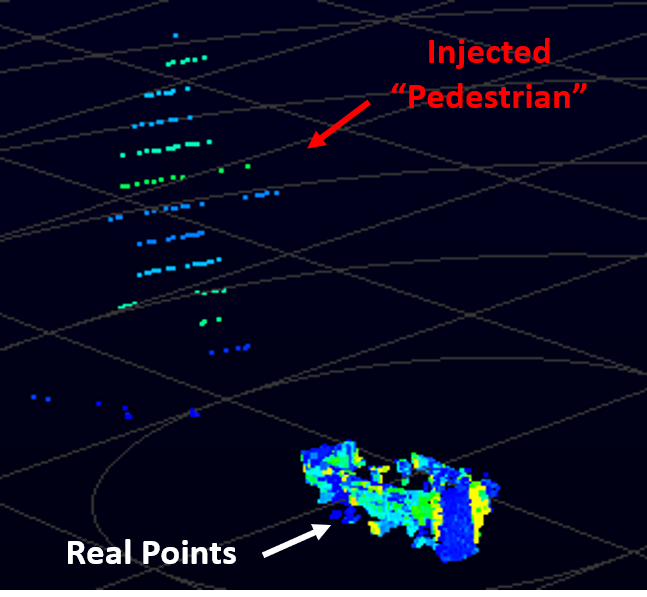}
		\label{fig:VLP-16_pedestrian}
	}	
	\caption{Points Injection with Different baseband signals. \textmd{(a) When the baseband signal is a periodic pulse signal, over 16000 controllable fake points can be injected. (b) With a fine-grained baseband signal, the pedestrian-pattern spoofing points can be injected.}}
		\vspace{-0.1in}
	\label{fig:impact_baseband}
\end{figure}

To inject as many points as possible, we designed the baseband signal to forge the echo signal in every receiving period of LiDAR. With the same baseband signal, we test the impact of different carrier signal frequencies to the number of injected points. We set the signal output power to 50 W and varied the carrier frequency between 700 MHz and 2500 MHz, recording the number of injected fake points. The results are shown in Fig.~\ref{fig:freq_vs_pointsnumber}. We find that different carrier frequencies indeed impact the number of injected points. We attribute this to the receiving circuit's varying response to different signal frequencies. Notably, a carrier frequency of approximately 1040 MHz enabled the injection of the highest number of points (over 16,500), forming a circular wall-like structure around the LiDAR, as shown in Fig.~\ref{fig:wall_injected}.At the same rotation speed, previous works could inject 3,000 fake points at most~\cite{jin2022pla,sato2023revisiting}. This significant increase is primarily because EM attacks affect a wider range (almost 360$^\circ$ horizontal angle), while laser attacks can only influence the area (less than 35$^\circ$ horizontal angle) illuminated by the laser spot.

 \subsubsection{Specified Pattern Injection}
Fig.~\ref{fig:impact_baseband} shows that different-pattern spoofing point cloud can be injected using various baseband signals.
With the fine-grained baseband signal and attack devices detailed in Sec.~\ref{sec:points_injection}, we successfully inject a "pedestrian" pattern into VLP-16 (Fig.~\ref{fig:VLP-16_pedestrian}) and RS-16 (Fig.~\ref{fig:synchronization}(c)in Appendix) LiDARs. This showcases that EM attacks possess capabilities akin to those of laser-based attacks~\cite{jin2022pla} for precisely manipulating the point cloud of a LiDAR system. 

\subsection{Feasibility Experiments on Moving Vehicle}
\label{sec:evaluation_moving}
In this section, we explore the feasibility of our attacks when the victim LiDAR is in motion. 
Videos of attacks can be found at our website~\cite{PhantomLiDAR}.


\subsubsection{Attack Setup}
\begin{figure}[pt]
    \centering

    \includegraphics[width=\linewidth]{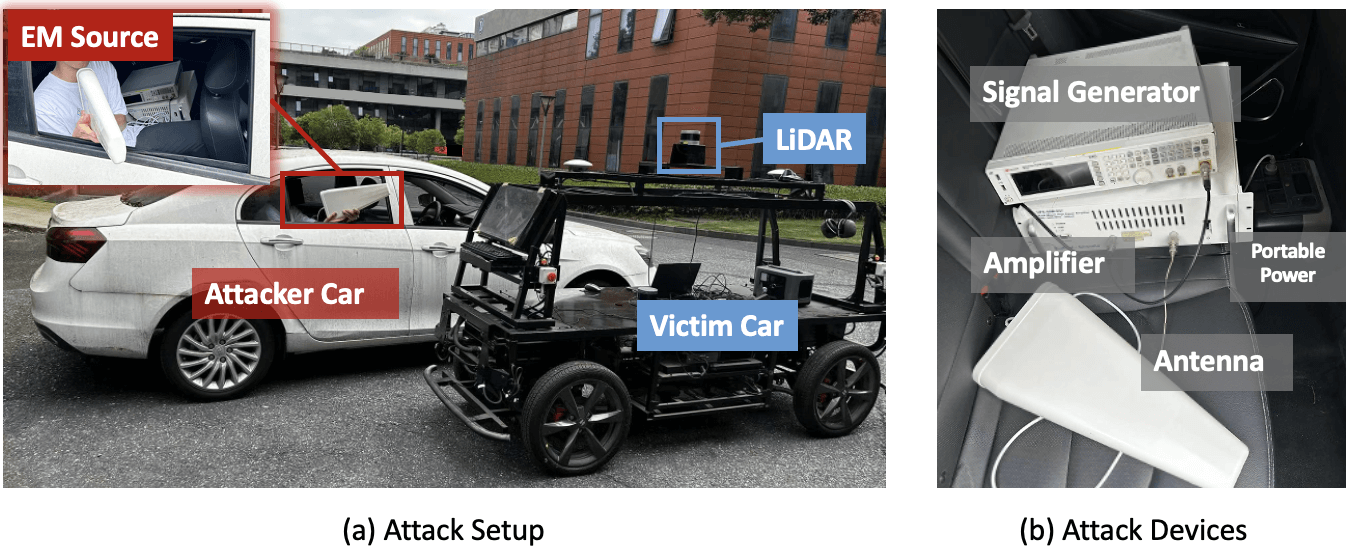}
    \vspace{-0.2in}
    \caption{Experimental Setup for the Attack on Moving Target.}
    \vspace{-10pt}
     \label{fig:moving_setup}
    
\end{figure}

The attack setting of moving vehicles is shown in Fig~\ref{fig:moving_setup}. The vehicle equipped with a VLP-16 LiDAR moves at a speed below 10 km/h (for safety considerations). During the experiment, we do not require the victim car to travel at a constant speed, which is in line with the real-world conditions where the attacker has no access to the victim car. 
The attack devices, including a signal generator, an amplifier, and an antenna, are integrated into the attack car. The attacker holds antenna in the car and aims it at the moving victim LiDAR. The attacker's goal is to compromise the victim LiDAR and make the LiDAR-based 3D object detection model unable to detect the attack car. Therefore, we set the parameters of attack signal to a frequency of 1.2GHz and an output amplitude of 50W, with which the \textit{Points Removal} can be easily achieved.


\subsubsection{Attack Scenarios}

We define two real-world attack scenarios. \textit{Scenario A} is the "tailgating attack", where both the attacker car and the victim car are moving on a straight road, and the driver of the attacker car will drive close to the victim car at a similar speed. \textit{Scenario B} is the "roadside attack", where the attacker is stationary at the roadside while the victim car is making a turn. Compared to the stationary attack in Sec.~\ref{sec:evaluation_point_eliminate}, \textit{Scenario A} primarily faces challenges due to the constantly changing attack distance and vehicle vibrations, while \textit{Scenario B} primarily faces aiming challenges due to the constantly changing attack angle.

\begin{figure}[tp]
	\centering
	\subfigure[Tailgating Attack]{
        \includegraphics[width=0.4\linewidth]{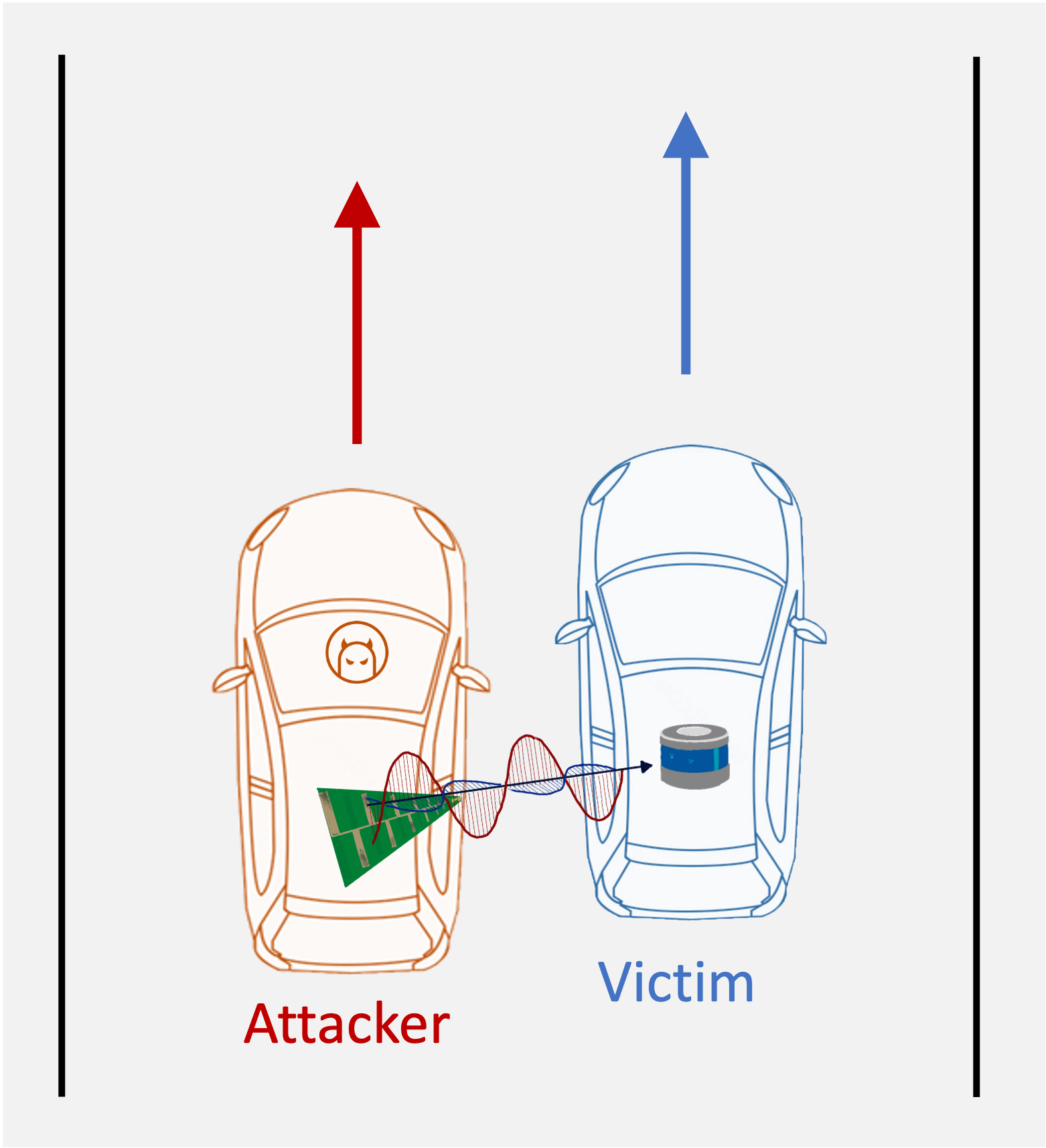}
         \label{fig:tailgating_attack}
	}
	\subfigure[Roadside Attack]{
	\includegraphics[width=0.4\linewidth]{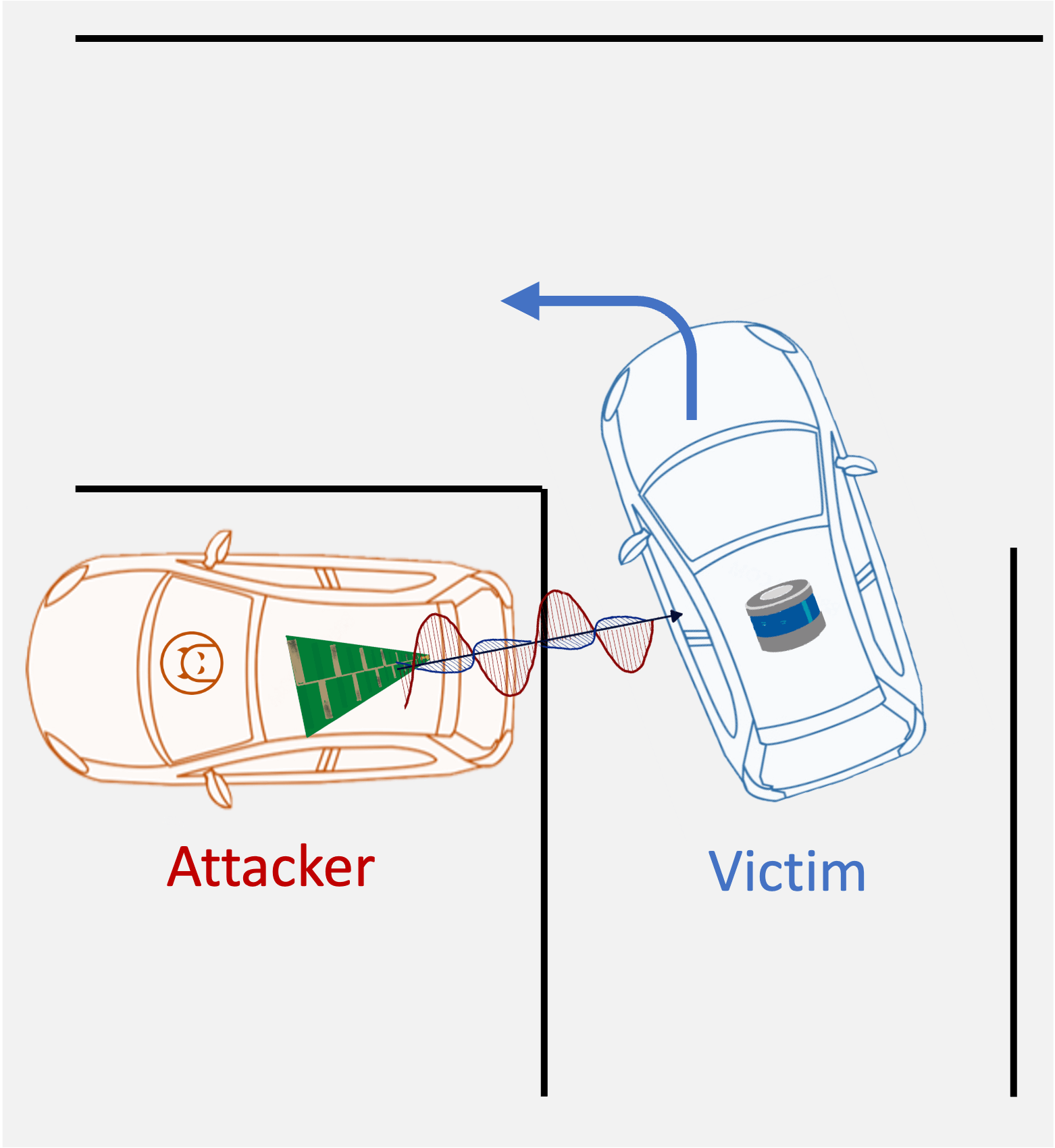}
         \label{fig:roadside_attack}
	}
        \vspace{-0.1in}
	\caption{Two Attack Scenarios. \textmd{(a) Tailgating attack: the attacker car drive close to the victim car at a similar speed. (b) Roadside attack: the attacker is stationary at the roadside while the victim car is making a turn.}}
        \label{fig:attack_scenarios_moving}
	\vspace{-0.1in}
\end{figure}

\subsubsection{Results}

We demonstrated the feasibility of hiding a specified target in both moving scenarios. 
Specifically, for the tailgating attack, we conducted five trials, each covering a distance of 40 meters. Each trial collected 100 frames uniformly, resulting in an attack success rate of 87.2\% (436/500). For the roadside attack, we also conducted five trials and collected 40 frames from each trial. We found that despite the continuously changing angle of the victim car, an attack success rate of 83.5\% (167/200) was still achievable within an attack distance of 4 meters.
In summary, due to the relatively low aiming requirement for EM attacks, as long as the attacker can approach the victim LiDAR (within 4 meters in this paper), the hiding attack can be carried out with above 80\% attack success rate.

\section{Discussion}

\begin{table}[tp]
\renewcommand\arraystretch{1.2}
 \centering
  \caption{Comparison with Laser-based Attacks}
  \label{tab:comparison}
  \vspace{-0.1in}
    \resizebox{\linewidth}{!}{\begin{tabular}{c|c|cccc|c|c}
\hline
\multirow{2}{*}{\textbf{Attack Type}}                     & \multirow{2}{*}{\textbf{Attack Surface}}                                                                 & \multicolumn{4}{c|}{\textbf{Attack Effect}}                                                                                                                                                                                                                                                                                                              & \multirow{2}{*}{\textbf{\begin{tabular}[c]{@{}c@{}}Aiming \\ Require.\end{tabular}}} & \multirow{2}{*}{\textbf{\begin{tabular}[c]{@{}c@{}}Attack \\ Distance\end{tabular}}} \\ \cline{3-6}
                                                          &                                                                                                          & \multicolumn{1}{c|}{\textbf{\begin{tabular}[c]{@{}c@{}}Points \\ Interference\end{tabular}}} & \multicolumn{1}{c|}{\textbf{\begin{tabular}[c]{@{}c@{}}Poionts \\ Removal\end{tabular}}} & \multicolumn{1}{c|}{\textbf{\begin{tabular}[c]{@{}c@{}}LiDAR \\ Poweroff\end{tabular}}} & \textbf{\begin{tabular}[c]{@{}c@{}}Points \\ Injection\end{tabular}} &                                                                                         &                                                                                      \\ \hline
Laser-based                                               & \begin{tabular}[c]{@{}c@{}}photodetector in \\ receiving circuit\end{tabular}                            & \multicolumn{1}{c|}{-}                                                                       & \multicolumn{1}{c|}{\begin{tabular}[c]{@{}c@{}}\small \checkmark \\~\cite{jin2022pla,cao2023you,sato2023lidar,suzukiwip} \end{tabular}  }                                                                   & \multicolumn{1}{c|}{-}                                                                  & \begin{tabular}[c]{@{}c@{}}\small \checkmark \\ ~\cite{petit2015remote,shin2017illusion,cao2019adversarialobject}\\~\cite{sun2020towards,jin2022pla,sato2023lidar}  \end{tabular}
                                                                   & High                                                                                    & >30meters                                                                             \\ \hline
\begin{tabular}[c]{@{}c@{}}EM-based\\ (Ours)\end{tabular} & \begin{tabular}[c]{@{}c@{}}receiving circuit, \\ monitoring sensor, \\ beam steering module\end{tabular} & \multicolumn{1}{c|}{\small \checkmark}                                                                       & \multicolumn{1}{c|}{\small \checkmark}                                                                   & \multicolumn{1}{c|}{\small \checkmark}                                                                  & \small \checkmark                                                                    & Low                                                                                     & 4 meters                                                                             \\ \hline
\end{tabular}}  


\end{table}

\subsection{Comparison with laser-based attack}

In this section, we compare EM-based attacks with recent laser-based attacks. It should be noted that the purpose of this comparison is not to determine whether EM-based or laser-based attacks are superior, as such a determination would require a more rigorous benchmark, including identical power levels or equivalent attack costs. Our aim is to provide the security community with a dialectical understanding of the characteristics of both types of attacks and to raise awareness, thereby inspiring more robust hardware design and customized defense strategies.

We compare attacks in terms of attack surfaces, attack effects, aiming requirements, and attack range, as shown in the Table~\ref{tab:comparison}. Overall, \alias exploits more attack surfaces than laser-based attacks to achieve a wider variety of attack effects and has the advantage of not requiring precise aiming. However, it is inferior to laser-based attacks in terms of attack distance.






\subsection{Feasibility Evaluation with Cheaper Hardware}
In the aforementioned experiments, we utilized high-end equipment to conduct wide-range frequency and amplitude sweeps, enabling us to identify new LiDAR vulnerabilities to IEMI signals. 
However, in practical scenarios, attackers could employ lower-cost equipment with a specific frequency signal generator and amplifier, which can also achieve the desired attack outcomes.
To further discuss the practicality of the attack in terms of cost, we conducted experiments using lower-cost equipment. Our first lower-cost setup, as shown in Fig.~\ref{fig:setup_160USD} included a 35MHz-4400MHz signal generator board(\$25~\cite{1.2GHz_Signal_Generator_25USD}) and a 1080MHz-1360MHz, 50W amplifier (\$128~\cite{1.2GHz_50W_128USD}). This \$168 atatck setup enabled us to achieve \textit{Points Interference}, \textit{Points Removal} and \textit{LiDAR Power-off} on the VLP-16 LiDAR.However, due to the inability of this setup to support amplitude modulation, \textit{Points Injection} was not feasible. 
Our second lower-cost setup, as shown in Fig.~\ref{fig:setup_2300USD}, included a USRP B210(\$2160~\cite{USRP_B210_2160USD}), a 50W amplifier(\$128) and an antenna (\$15). Using this \$2300 setup where the USRP B210 supports signal amplitude modulation, we successfully conducted all four types of attack effects, including \textit{Points Injection} attack. 


\subsection{Countermeasures}
When utilizing LiDAR in safety-critical scenarios, countermeasures against \textit{PhantomLiDAR} attacks can be implemented through the following approaches:

\textbf{EMC Reinforcement:}
The attack interfaces of \textit{PhantomLiDAR} are the analog portion of the LiDAR. There are three common defenses ~\cite{kune2013ghost} to reinforce the EMC of analog circuits: shielding, differential comparators, and filters. Shielding involves the application of a conducting material to shield a component from EMI. Where shielding is either not possible or not sufficient, a reference signal can be used to remove the common mode voltage using a differential circuit~\cite{razavi2005design}. Additionally, a filter that attenuates signals outside a sensor’s baseband frequency can reduce the vulnerable frequency range of that sensor.
However, these methods can increase the complexity and cost of LiDAR circuits. 
While COTS LiDAR systems undergo rigorous EMC testing before shipment, the experiments in this paper demonstrate that adversaries can easily compromise LiDAR with commercial devices. Therefore, it may be prudent to consider updating the EMC standards for LiDAR systems.


\textbf{Multi-Sensor Fusion:}
This paper demonstrates through experimentation in Sec.~\ref{sec:evaluation_point_interference} that fusion models show promise in mitigating the effects of the attack. In terms of autonomous vehicles, perception can be enhanced by the fusion of camera and LiDAR, and by equipping multiple LiDARs to increase safety redundancy.  Therefore, how multiple sensors can better cooperate and complement each other presents an intriguing research question worth exploring~\cite{jin2024unity}.

\subsection{Future Work}
Although this paper extensively evaluates the \textit{Phantom LiDAR} attack in a physical world setting, it has not been tested on commercial autonomous vehicles equipped with LiDAR. Such an effort may require authorization and collaboration with automotive manufacturers. Due to the integration of multiple sensors and the presence of multiple safety assurance measures~\cite{iso26262} in commercial vehicles, we believe that end-to-end testing of PhantomLiDAR on autonomous vehicles could yield novel and interesting effects, which we will investigate in future research.

\subsection{Responsible Disclosure}
We have disclosed the EM vulnerability discovered in this work to the relevant product security teams~\cite{Ouster,Robosense}, and provided the experimental steps to reproduce the attacks. In addition, we suggested some potential methods to mitigate our attacks.


\section{Related Work}

\subsection{LiDAR Security}

The output of LiDAR (i.e. point cloud) can be manipulated by laser and 3D object. 
Based on the inherent vulnerability of photoelectric sensors, saturation attack against LiDAR can be easily induced by high-power continuous laser~\cite{shin2017illusion}. After the iterative efforts of several papers~\cite{petit2015remote,shin2017illusion,cao2019adversarial,sun2020towards}, the latest literature~\cite{jin2022pla} proves that a large number of controllable points can be injected into the mechanical (spinning) LiDAR through carefully designed laser pulses, and physically validates the feasibility of hiding attack and creating attack.  However, since lasers share the same physical channel as LiDAR, the ability of lasers to compromise LiDAR systems is an unavoidable phenomenon. This makes it challenging to propose improvements to the safety design of LiDAR systems that would fundamentally prevent laser attacks. Additionally, since laser attacks generally face aiming concerns, the real-world threats of laser attacks has not been widely recognized.
 The way of using 3D objects to manipulate point clouds is also very popular. There are mainly two methods: 3D printing objects and placing arbitrary objects. 3D printing object can realize the adversarial point cloud of specific shape in the physical world, and make the attack difficult to be aware by human beings while spoofing the victim detection model~\cite{cao2019adversarialobject,tu2020physically,yang2021robust}. Some studies have found that for the adversarial attack that aims to hide a target, the position of the adversarial points is more critical than the shape, thus the adversarial effect can be realized by placing the arbitrary object at the specified position~\cite{zhu2021can,zhu2021adversarial}. A recent work~\cite{bhupathiraju2023emi}  demonstrates that EMI can compromise LiDAR’s ToF circuits and induce attack effecrs of "sensor data perturbations". Compared to this work, 
 \alias demonstrates 3 new attack effects with new attack surfaces and signal design methods.
 

\subsection{IEMI Attacks on Cyber-physical Systems}
Intentional electromagnetic interference (IEMI) attacks can induce parasitic current or voltage into the target cyber-physical systems to manipulate their output data or gain unauthorized access~\cite{tu2019trick,liu2023magbackdoor,kune2013ghost,kasmi2015iemi,dai2023inducing,liu2023magbackdoor,esteves2018remote,fokkens2023prediction,maruyama2019tap,wang2022ghosttouch,shan2022invisible,gao2023expelliarmus,jiang2024ghosttype,mohammed2022towards,jiang2023glitchhiker,bhupathiraju2023emi,jang2023paralyzing,pahl2021in,pahl2022analysis,selvaraj2018electromagnetic,dayanikli2022physical,dayanikli2021electromagnetic} and have been of vital interest in recent years.
For example, existing works use IEMI attacks to change the output of microphones~\cite{kune2013ghost,kasmi2015iemi,dai2023inducing,liu2023magbackdoor,esteves2018remote,fokkens2023prediction}, touchscreen~\cite{maruyama2019tap,wang2022ghosttouch,shan2022invisible,gao2023expelliarmus}, keyboard~\cite{jiang2024ghosttype} and smart lock~\cite{mohammed2022towards} to allow attackers to inject false speech audios and user inputs, often to pretend that a real user is interacting with the smartphones or computer systems.
Besides, IEMI attacks are also frequently used to change how autonomous systems perceive their surrounding environment, e.g., by changing the output of camera~\cite{jiang2023glitchhiker}  and inertial measurement units~\cite{jang2023paralyzing,pahl2021in,pahl2022analysis,dayanikli2022physical} on autonomous vehicles.

\section{Conclusion}

In this paper,  we uncovered and experimentally validated new vulnerabilities in LiDAR systems, including novel attack surfaces and attack causalities. We validated that the receiving circuit, monitoring sensors(temperature sensor), and optical encoder in beam-steering module in LiDAR can serve as EMI attack surfaces. We identified two primary causalities: direct interference with ranging due to EM coupling into the receiving circuit, and indirect compromise of LiDAR by exploiting its fault management mechanisms.
Based on the new attack surfaces and attack principles, we introduce four types of EM-based attacks against LiDAR. To the best of our knowledge, we are the first to design and propose \textit{Points Removal}, \textit{LiDAR Power-off}, and \textit{Points Injection} attacks with EMI. Additionally, compared to prior SOTA works, our attack capabilities show significant improvements in terms of the ranging errors (2x more) and the number of fake points (5x more). 
Comprehensive experiments conducted on five LiDARs and five models demonstrate the efficacy of our attacks. 
The attack's practical threat is evidenced by its considerable attack distances, its low aiming requirements, and its feasibility in moving scenarios. In addition, we discussed the countermeasures proposed against our attacks.
We hope our research can enhance future LiDAR systems by considering a wider range of attack vectors. Future directions include exploring the feasibility of PhantomLiDAR on autonomous vehicles.






\bibliographystyle{IEEETranS.bst}
\bibliography{reference.bib}



\appendix
\section{Appendix}

\subsection{The Fault Diagnostic and Management}

In the realm of electronic device engineering, The Fault Diagnostic and Management (FDM) mechanisms play a pivotal role in ensuring the reliability and efficiency of electronic devices. FDM methodologies are designed to detect, diagnose, and manage faults within electronic circuits and components, minimizing downtime and preventing catastrophic failures. These systems leverage a combination of advanced algorithms, including machine learning and artificial intelligence, to analyze data from sensors and other sources. This data-driven approach enables the early detection of anomalies, facilitating prompt intervention. Furthermore, FDM mechanisms are integral in predicting potential failures through proactive health monitoring, thereby extending the lifespan of electronic components. The integration of FDM mechanisms in electronic devices not only enhances performance but also contributes significantly to safety, particularly in critical applications such as medical devices, automotive electronics, and aerospace systems. As technology advances, the complexity and sophistication of FDM mechanisms continue to evolve, offering more robust and intelligent solutions for fault management in the ever-expanding landscape of electronic devices.

\subsection{The Necessity of Synchronization for Point Control}
\begin{figure}[t]
    \centering

    \includegraphics[width=0.48\textwidth]{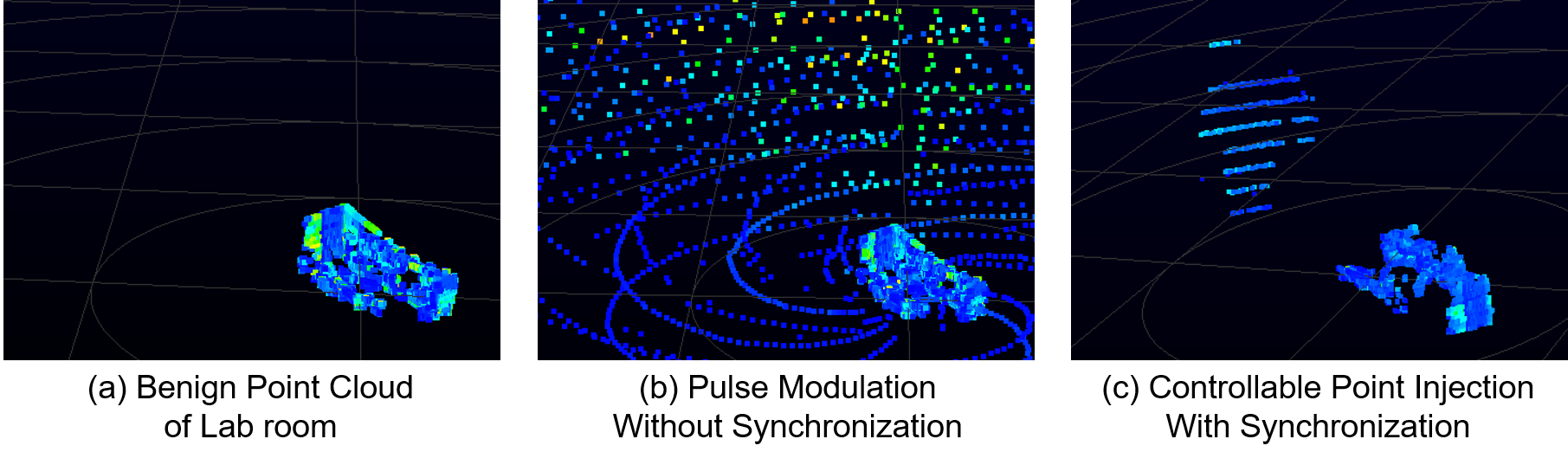}
    \vspace{-0.2in}
    \caption{Illustration of the necessity of synchronization for point control. 
    \textmd{(a)Benign point cloud of the lab room. (b) Random fake points injected by pulse-modulated EM signals without synchronization. (c) Fake "pedestrian" points injected by modulated EM signals with synchronization.}
    }
    \label{fig:illustration_necessity_synchornization}
    \vspace{-10pt}
    \label{fig:synchronization}
\end{figure}

 The necessity of synchronization is illustrated in Fig.~\ref{fig:synchronization}. Without a synchronization mechanism based on receiving and delaying, as shown in Fig.~\ref{fig:synchronization}(b), the injected point cloud would be random. 
In contrast, as shown in Fig.\ref{fig:synchronization}(c), when synchronization is induced, the desired fake points can be injected.
It is worth noting that, for a more intuitive demonstration of the effect without synchronization, the signal used in Fig.\ref{fig:synchronization}(b) is designed with a pulse at every receiving time of the LiDAR. If the signal from Fig.\ref{fig:synchronization}(c) is used without synchronization, the scarcity of pulses may often result in the inability to inject a point cloud.



\begin{figure}[tp]
	\centering

 
        \includegraphics[width=0.95\linewidth]{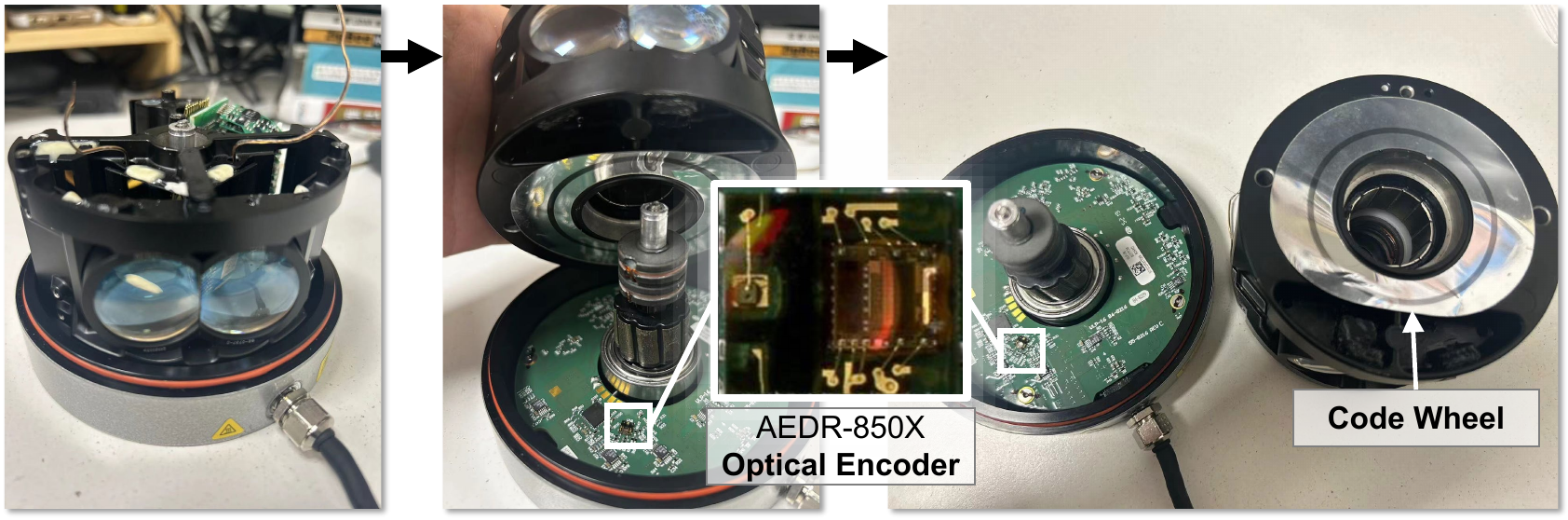}

        

	\caption{The Teardown of VLP-16 LiDAR. \textmd{By disassembling the LiDAR, we find that the VLP-16's bottom board is equipped with an optical encoder, AEDR-850X, utilized for measuring the rotational speed of the LiDAR.}}
        \label{fig:Optical_Encoder_Teardown}
\end{figure}

\begin{table}[]
\renewcommand\arraystretch{1.2}
 \centering
  \caption{Average Precision (AP) of Five 3D Object Detection Models under Different-Intensity Points Interference}
  \label{tab:interfernce_result}
\resizebox{1\linewidth}{!}{
\begin{tabular}{|c|cc|ccc|}
\hline
\toprule
\multirow{2}{*}{\textbf{\begin{tabular}[c]{@{}c@{}}Interference \\ Intensity\end{tabular}}} & \multicolumn{2}{c|}{\textbf{LiDAR-only}}                                          & \multicolumn{3}{c|}{\textbf{LiDAR-Camera Fusion}}                                                                   \\ \cline{2-6} 
                                                                                            & \multicolumn{1}{l|}{\textbf{PointPillar}} & \multicolumn{1}{l|}{\textbf{PV-RCNN}} & \multicolumn{1}{l|}{\textbf{VirConv-L}} & \multicolumn{1}{l|}{\textbf{EPNet}} & \multicolumn{1}{l|}{\textbf{CLOCs}} \\ \hline
0                                                                                           & \multicolumn{1}{c|}{77.616}               & 83.660                                & \multicolumn{1}{c|}{86.818}             & \multicolumn{1}{c|}{78.831}         & 76.885                              \\ \hline
5cm                                                                                         & \multicolumn{1}{c|}{77.068}               & 79.101                                & \multicolumn{1}{c|}{86.006}             & \multicolumn{1}{c|}{78.273}         & 76.267                              \\ \hline
10cm                                                                                        & \multicolumn{1}{c|}{72.585}               & 77.069                                & \multicolumn{1}{c|}{85.651}             & \multicolumn{1}{c|}{76.746}         & 71.918                              \\ \bottomrule \hline
\end{tabular}}
\end{table}

\begin{figure}[t]
    \centering
    \includegraphics[width=0.45\textwidth]{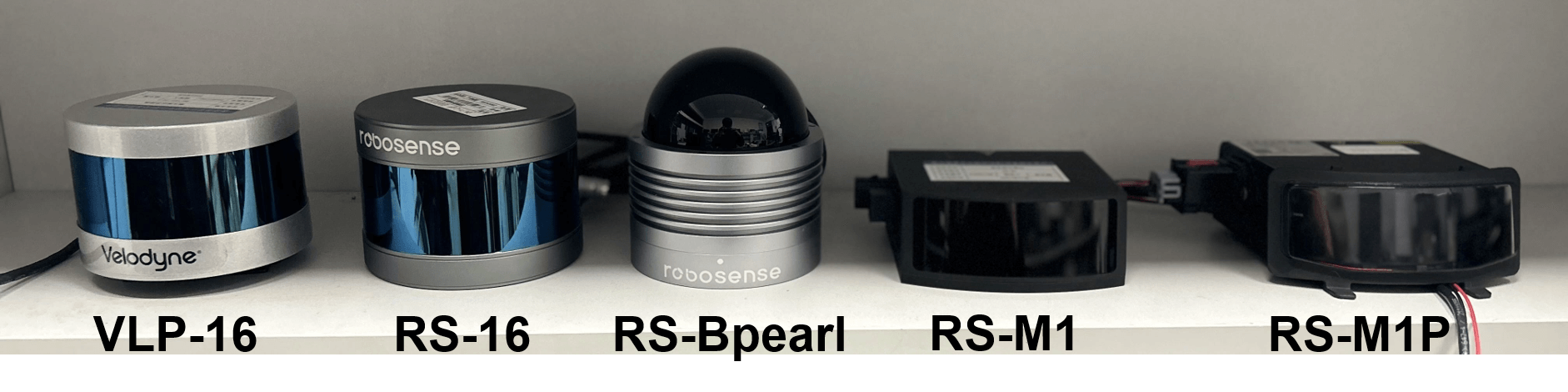}
    \vspace{-0.1in}
    \caption{The LiDARs under test.}
    \vspace{-10pt}
     \label{fig:LiDARs}
    
\end{figure}

\begin{figure}[t]
    \centering

    \includegraphics[width=0.45\textwidth]{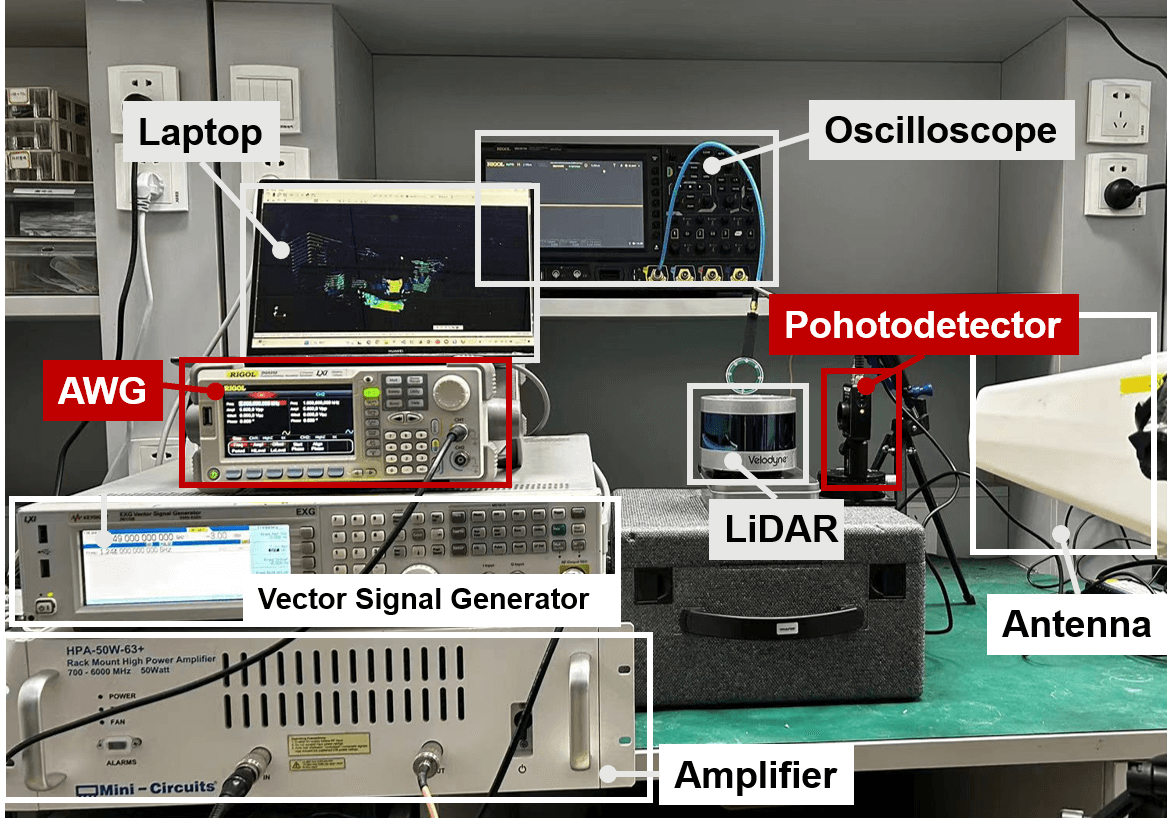}
    \vspace{-0.1in}
    \caption{Point Injection Setup. \textmd{Compared to the setup of the other 3 types of attacks, an arbitray signal generator and a photodetector are added.}}
    \vspace{-10pt}
     \label{fig:Point_Injection_Setup}
    
\end{figure}

\begin{figure}[t]
	\centering
        \subfigure[The Attack Setup Costs \$160]{
		\includegraphics[width=0.45\linewidth]{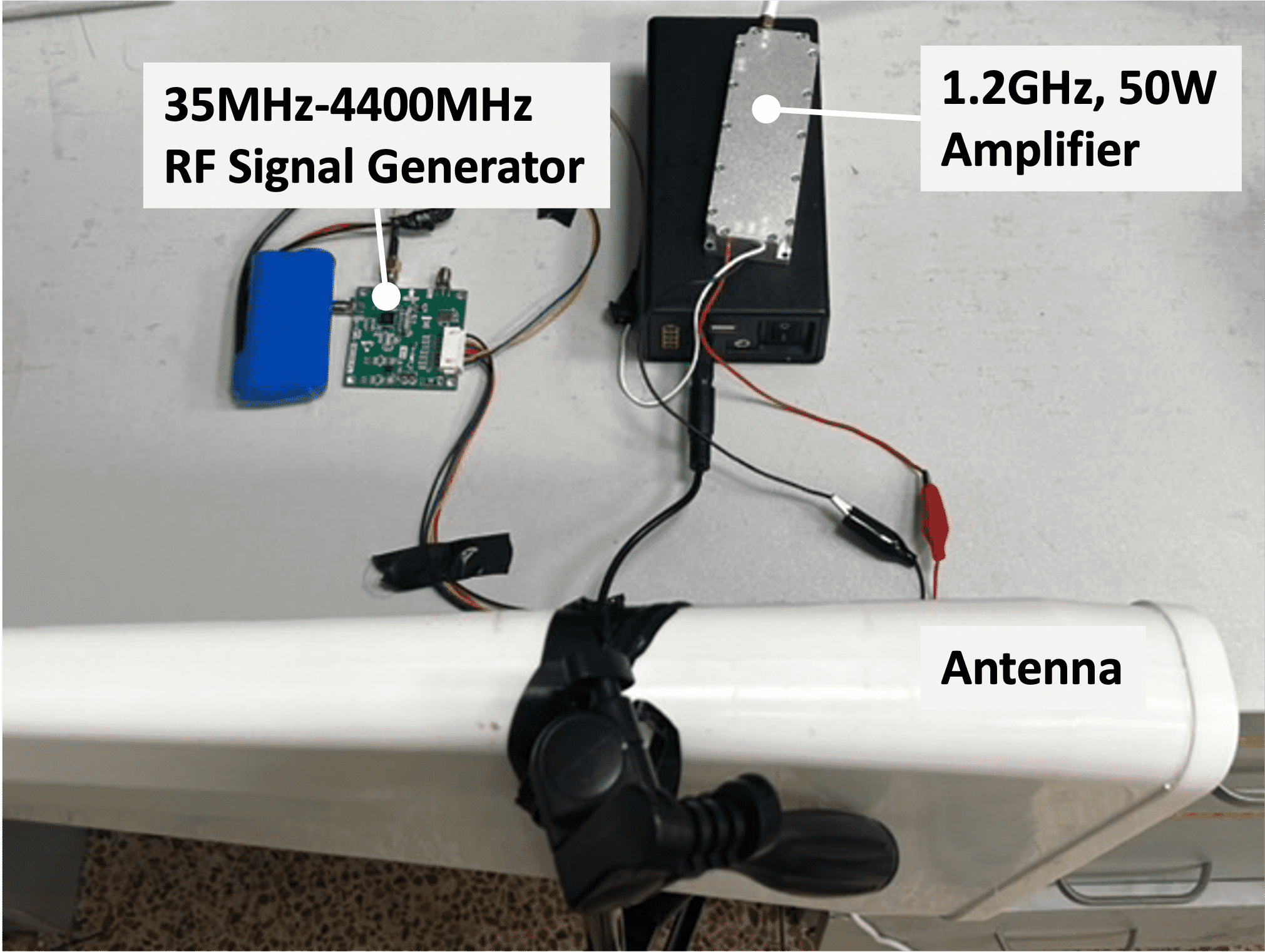}
		\label{fig:setup_160USD}
	}	
	\subfigure[The Attack Setup Costs \$2300]{
		\includegraphics[width=0.45\linewidth]{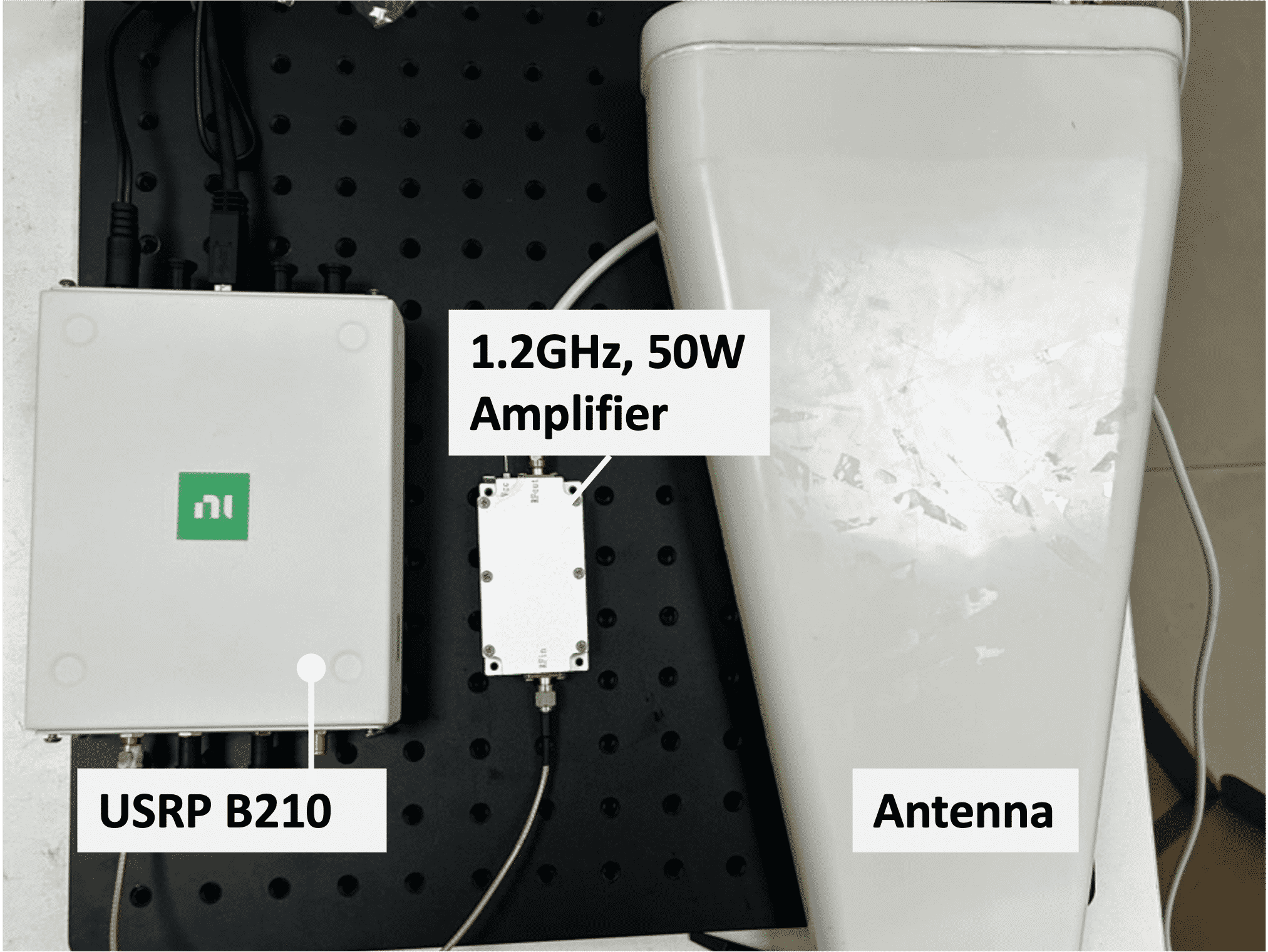}
		\label{fig:setup_2300USD}
	}	
	\caption{Attack Setup with Cheaper Hardwares. \textmd{With approximately \$168 worth of equipment, we were able to achieve \textit{Points Interference}, \textit{Points Removal} and \textit{LiDAR Power-off}. With approximately \$2300 worth of equipment, we were able to achieve \textit{Points Injection}.}}
		\vspace{-0.1in}
	\label{fig:cheaper_setup}
\end{figure}


\begin{table*}[h]
    \centering
    
    \caption{Fault Detection and Diagnostic List from an Anonymous LiDAR Manufacturer}
    \begin{tabular}{c}
    \hline
    \includegraphics[width=1\linewidth]{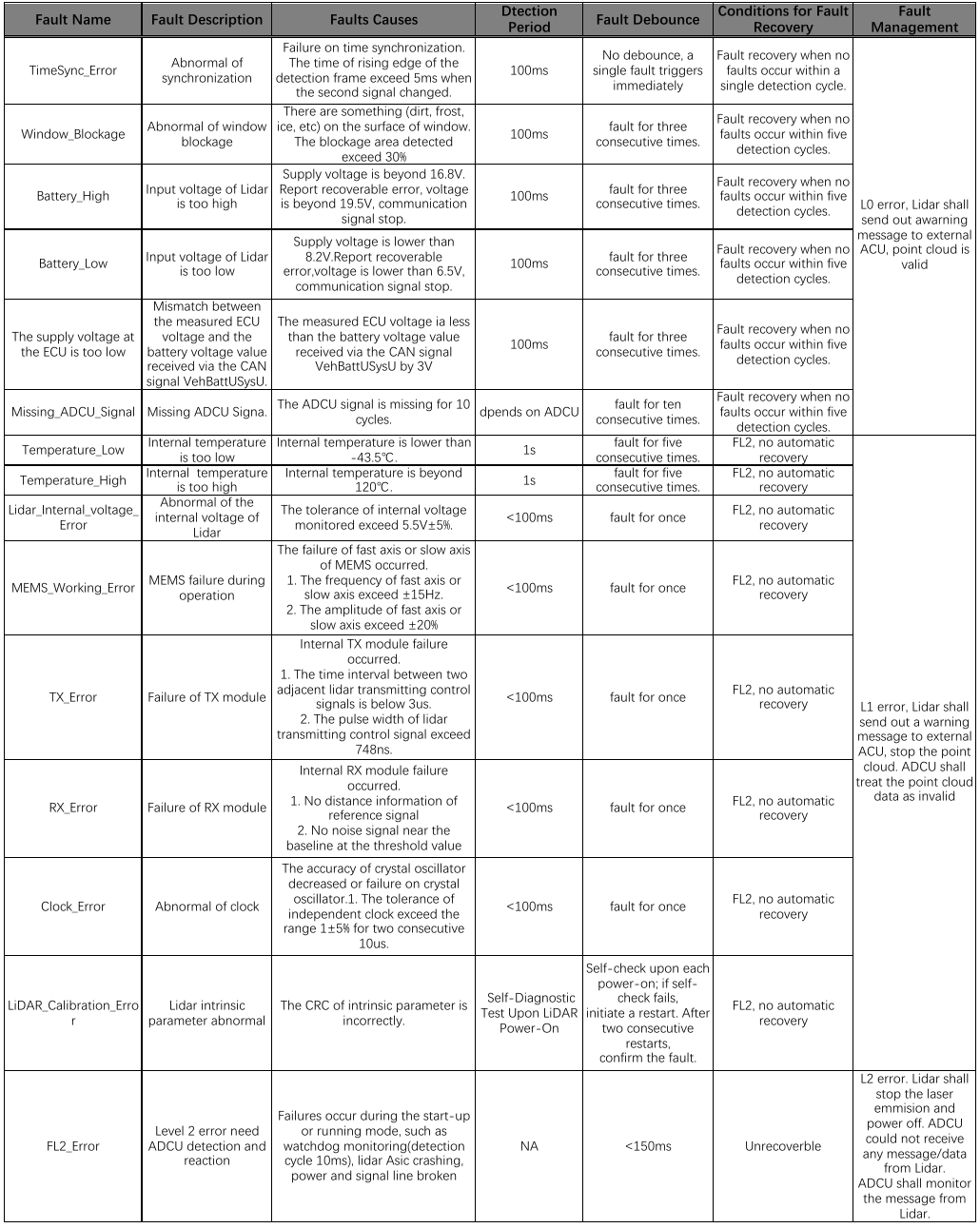}\\
    \hline
    \end{tabular}
    \label{table:Fault_Matrix}
\end{table*}

\end{document}